\documentclass[twocolumn]{aastex631} 
\PassOptionsToPackage{svgnames}{xcolor}
\usepackage{graphicx}	
\usepackage{amsmath}	
\usepackage{amssymb}	
\usepackage{bm}		
\usepackage{float}
\usepackage{natbib}
\usepackage{scalerel}
\usepackage{comment}
\usepackage{hyperref}
\usepackage{changepage}
\usepackage{color,soul}

\newcommand{\agel}{{\tt AGEL}}
\newcommand{\glee}{\textsc{GLEE}}

\newcommand{\astrod}{{\tt ASTRO\hspace{0.4ex}3D}}

\newcommand{\sdss}{{\tt SDSS}}
\newcommand{\boss}{{\tt BOSS}}

\newcommand{\des}{{\tt DES}}

\newcommand{\decals}{{\tt DECaLS}}
\begin{document}
\shorttitle{gamma-redshift}
\shortauthors{Sahu et al.}
\defcitealias{Tran:Harshan:2022}{AGEL-DR1}
\received{February 3, 2024}
\revised{April 30, 2024}
\accepted{May 16, 2024, for publication in the Astrophysical Journal}

\maxdeadcycles=100000

\title{AGEL: Is the Conflict Real? Investigating Galaxy Evolution Models using Strong Lensing at $0.3 < z < 0.9$}

\correspondingauthor{Nandini Sahu}
\email{n.sahu@unsw.edu.au}

\author[0000-0003-0234-6585]{Nandini Sahu}
\affil{University of New South Wales, Sydney, NSW 2052, Australia}
\affil{The ARC Centre of Excellence for All Sky Astrophysics in 3 Dimensions (ASTRO 3D), Australia}

\author[0000-0001-9208-2143]{Kim-Vy Tran}
\affil{Center for Astrophysics, Harvard \& Smithsonian, Cambridge, MA 02138, USA}
\affil{The ARC Centre of Excellence for All Sky Astrophysics in 3 Dimensions (ASTRO 3D), Australia}
\affil{University of New South Wales, Sydney, NSW, Australia}

\author{Sherry H.~Suyu}
\affil{Technical University of Munich, TUM School of Natural Sciences, Department of Physics,  James-Franck-Stra{\ss}e 1, 85748 Garching, Germany}
\affil{Max-Planck-Institut f{\"u}r Astrophysik, Karl-Schwarzschild Stra{\ss}e 1, 85748 Garching, Germany}
\affil{Academia Sinica Institute of Astronomy and Astrophysics (ASIAA), 11F of ASMAB, No.1, Section 4, Roosevelt Road, Taipei 10617, Taiwan}

\author{Anowar J.~Shajib}
\affil{Department of Astronomy \& Astrophysics, The University of Chicago, Chicago, IL 60637, USA}
\affil{Kavli Institute for Cosmological Physics,
    University of Chicago, Chicago, IL 60637, USA}
\affil{NHFP Einstein Fellow}

\author{Sebastian Ertl}
\affil{Technical University of Munich, TUM School of Natural Sciences, Department of Physics,  James-Franck-Stra{\ss}e 1, 85748 Garching, Germany}
\affil{Max-Planck-Institut f{\"u}r Astrophysik, Karl-Schwarzschild Stra{\ss}e 1, 85748 Garching, Germany}

\author{Glenn G. Kacprzak}
\affil{Centre for Astrophysics and Supercomputing, Swinburne University of Technology, Hawthorn, Victoria 3122, Australia}
\affil{The ARC Centre of Excellence for All Sky Astrophysics in 3 Dimensions (ASTRO 3D), Australia} 

\author{Karl Glazebrook}
\affil{Centre for Astrophysics and Supercomputing, Swinburne University of Technology, Hawthorn, Victoria 3122, Australia}
\affil{The ARC Centre of Excellence for All Sky Astrophysics in 3 Dimensions (ASTRO 3D), Australia}

\author[0000-0001-5860-3419]{Tucker Jones}
\affil{Department of Physics and Astronomy, University of California, Davis, 1 Shields Avenue, Davis, CA 95616, USA}

\author[0000-0002-2645-679X]{Keerthi Vasan G.C.}
\affil{Department of Physics and Astronomy, University of California, Davis, 1 Shields Avenue, Davis, CA 95616, USA}

\author{Tania M. Barone}
\affil{University of New South Wales, Sydney, NSW, Australia}
\affil{Centre for Astrophysics and Supercomputing, Swinburne University of Technology, Hawthorn, Victoria 3122, Australia}
\affil{The ARC Centre of Excellence for All Sky Astrophysics in 3 Dimensions (ASTRO 3D), Australia}

\author{A. Makai Baker}
\affil{School of Physics and Astronomy, Monash University, Clayton VIC 3800, Australia}

\author{Hannah Skobe}
\affil{Department of Astronomy \& Astrophysics, The University of Chicago, Chicago, IL 60637, USA}

\author{Caro Derkenne}
\affil{Astrophysics and Space Technologies Research Centre, School of Mathematical and Physical Sciences, Macquarie University, NSW 2109, Australia}
\affil{The ARC Centre of Excellence for All Sky Astrophysics in 3 Dimensions (ASTRO 3D), Australia}

\author[0000-0003-3081-9319]{Geraint F. Lewis}
\affil{Sydney Institute for Astronomy, School of Physics A28, The University of Sydney, NSW 2006, Australia}

\author[0000-0002-1576-2505]{Sarah M. Sweet}
\affil{School of Mathematics and Physics, University of Queensland, Brisbane, QLD 4072, Australia}
\affil{The ARC Centre of Excellence for All Sky Astrophysics in 3 Dimensions (ASTRO 3D), Australia}

\author{Sebastian Lopez}
\affil{Departamento de Astronomía, Universidad de Chile, Camino el Observatorio 1515, Las Condes, Santiago, Chile}

\nocollaboration{16}

\newcommand{\zdefl}{$z_{\rm defl}$}
\newcommand{\zsrc}{$z_{\rm src}$}


\begin{abstract}
Observed evolution of the total mass distribution with redshift is crucial to testing galaxy evolution theories. To measure the total mass distribution, strong gravitational lenses complement the resolved dynamical observations currently limited to $z\lesssim 0.5$. Here we present the lens models for a pilot sample of seven galaxy-scale lenses from the \astrod\ Galaxy Evolution with Lenses (\agel) survey. The \agel\ lenses, modeled using HST/WFC3-F140W images with Gravitational Lens Efficient Explorer (\glee) software, have deflector redshifts between $0.3<$\zdefl$<0.9$. Assuming a power-law density profile with slope $\gamma$, we measure the total density profile for the deflector galaxies via lens modeling. We also measure the stellar velocity dispersions ($\sigma_{\rm obs}$) for four lenses and obtain $\sigma_{\rm obs}$ from \sdss-\boss\ for the remaining lenses to test our lens models by comparing observed and model-predicted velocity dispersions. For the seven \agel\ lenses, we measure an average density profile slope of $-1.95\pm0.09$ and a $\gamma$--$z$ relation that does not evolve with redshift at $z<1$. Although our result is consistent with some observations and simulations, it differs from other studies at $z<1$ that suggest the $\gamma$--$z$ relation evolves with redshift. The apparent conflicts among observations and simulations may be due to a combination of 1) systematics in the lensing and dynamical modeling; 2) challenges in comparing observations with simulations; and 3) assuming a simple power-law for the total mass distribution. By providing more lenses at \zdefl$>0.5$, the \agel\ survey will provide stronger constraints on whether the mass profiles evolve with redshift as predicted by current theoretical models.

\end{abstract}

\section{Introduction}
\label{Sec:Introduction}

In standard cosmology, galaxies form via baryonic and dark matter assembly in the overdense centers of cold dark matter halos \citep{Blumenthal:Faber:1984, White:Frenk:1991}. 
Further, galaxies grow their mass and morphology via mergers and environmental processes \citep{Somerville:Dave:2015}.
To test galaxy evolution theories and various elements adopted in simulations, such as the cosmological model, dark matter properties and its dynamics with baryons, baryonic feedback, and subgrid physics, comparisons of observed evolution of total mass distribution with cosmological simulations are essential.

To measure the total baryonic plus dark matter halo mass distribution, gravitational lensing is a powerful tool from galactic (kpc) to cluster (Mpc) scales \citep[e.g., see][]{Blandford:Narayan:1992,Treu:2010,Shajib:Vernardos:Collett:2022}.
Gravitational lensing only depends on the total mass distribution of the deflector (also called the lens),  and it is independent of the deflector's luminosity or composition.  
Galaxy-scale strong lenses, where the deflector is a single galaxy, can be used to measure the total mass distribution of the deflector galaxy at high redshifts \citep[e.g.,][]{Sonnenfeld:Gavazzi:2013}, which are otherwise measured using resolved kinematic observations currently limited to lower redshifts \citep[$z < 0.5 $,][]{Derkenne:McDermid:Poci:2023}.

The mass distribution of deflectors in galaxy-scale lenses is commonly represented by a radial power-law density profile, $\rho \propto r^{\gamma}$, with a constant effective slope $\gamma <0$ \citep{Treu:Koopmans:2002a}. 
The observed evolution of density profile slope with redshift, i.e., the $\gamma$--$z$ relation, is compared with the predictions from simulations to examine the galaxy evolution theories.
This investigation has been limited to early-type galaxies (ETGs) because, in the current observational settings, the deflectors in the observed galaxy-scale lens samples are commonly massive ETGs  \citep[e.g.,][]{2006ApJ...638..703B}.

Massive early-type galaxies are theorized to evolve broadly in two phases \citep{Naab:Johansson:2007, Guo:White:2008, Oser:Ostriker:2010}. 
In the first phase, at high redshift ($z \gtrsim 2$),  growth happens through gas-rich accretion and mergers. Here abundant cold gases move towards the inner galaxy potential well and allow dominantly in-situ star formation, increasing the inner galaxy mass more than the outer galaxy regions. 
Thus, in this phase, the total mass density profile slope, $\gamma$, steepens as the redshift decreases (i.e., $d\langle \gamma \rangle/dz > 0$). 

In the second phase, at lower redshifts ($z \lesssim 1-2$), as the cool gas is exhausted and in-situ star formation ceases, the dominant mode of evolution is mass assembly via gas-poor (dry) mergers and dry accretion \citep{Oser:Ostriker:2010}. 
Major dry mergers puff up the galaxy with a considerable increase in mass and size, resulting in a shallower radial density profile slope. 
Thus, $\gamma$ shallows as redshift decreases (i.e., $d\langle \gamma \rangle/dz < 0$), such that the density profile tends to be isothermal \citep[i.e., $\gamma=-2$,][]{Blandford:Kochanek:1987} as $z \rightarrow 0 $.
On the other hand, in the same period, galaxies evolving via gas-rich processes tend to maintain a steeper density profile as the redshift decreases \citep{Barnes:Hernquist:1991, Mihos:Hernquist:1994}.

Current advanced cosmological simulations, e.g., Magneticum \citep{Remus:Dolag:Naab:2017} and IllustrisTNG \citep{Wang:Vogelsberger:2019}, which also incorporate stellar and active galactic nuclei (AGN) feedback models are broadly  consistent with the above galaxy evolution model. 
Especially, Magneticum and IllustrisTNG simulations predict that from $z=2 \ \rm to \ 0$, the density profile slope of ETGs shallows, starting from $\gamma$ between $-2.2$ to $-3.5$ at $z\sim 2$ to $\gamma=-2$ as  $z \rightarrow 0$.

Observational probes such as the dynamics of HI gas, globular clusters, planetary nebulae, and analysis of X-ray gas temperatures have found an average isothermal slope ($\gamma \sim -2 $) for local ($z\sim 0$) ETGs \citep{Weijmans:Krajnovic:2008, Brodie:Romanowsky:2014, Coccato:Gerhard:2009, Humphrey:Buote:2010}. 
Using the new 3D integral field spectroscopic observations with high-resolution 2D kinematics, \citet{Poci:Cappellari:2017} and \citet{Derkenne:McDermid:2021, Derkenne:McDermid:Poci:2023} also found an average isothermal density profile for ETGs and suggested that $\gamma$ does not evolve with redshift, i.e. $d\langle \gamma \rangle/dz \sim 0$ at $0<z\lesssim 0.5$. 

Most lensing observations so far have either suggested a slightly steeper slope than an isothermal density profile  \citep[$\gamma \sim -2.1$,][]{Auger:Treu:Bolton:2010,  Barnabe:Czoske:2011} or a gradual steepening of the slope with decreasing redshift  \citep[e.g.,][]{Bolton:Brownstein:2012, Li:Shu:Wang:2018} for deflector ETGs. 
Therefore, regarding the evolution of density profile slope with redshift, lensing observations are apparently at conflict with simulations and marginally so with purely kinematic observations described in above paragraphs. 
However, the $z \gtrsim 0.5$ end of the observed $\gamma$--$z$ relation is still too sparsely populated to suggest a robust $\gamma$--$z$ trend and requires a larger lens sample.

If true, the interpretation of most lensing observations so far suggests that massive ETGs instead require a gas-rich mass growth model, even at $z<1$ \citep{Sonnenfeld:Nipoti:2014}. 
However, it is contrary to the predictions from ETG 
evolution theories.
A few studies have tried to address this problem. \citet{Remus:Dolag:Naab:2017} argue that the mismatch between simulations and lensing observations is artificial due to incorrect comparison. 
 \citet{Etherington:Nightingale:2023}, however, suggest a mismatch among lensing observations due to the use of different lens modeling methods, ``lensing-only" or  ``lensing and dynamical" (L\&D), to measure the  density profile slope.
Thus, whether or not the conflict is real requires further detailed analysis.

This paper aims at populating the observed $\gamma$--$z$ relation, especially at higher redshift ($\gtrsim 0.5$), using deflector galaxies of seven galaxy-scale lenses from the \astrod \, Galaxy Evolution with Lenses (\agel) survey \citep[][hereafter, \agel-DR1]{Tran:Harshan:2022}. 
Here we also try to understand the $\gamma$--$z$ relation by comparing lensing observations, dynamical observations, and simulations. 

The deflectors in our pilot lens sample of seven are all  massive ETGs with redshifts in the range $0.3 < z < 0.9$, five of these galaxies are at $z \gtrsim 0.5$.
To obtain the mass density profile of deflectors, we model \agel \, lenses using the ``Lensing-only" method with the advanced lens modeling software Gravitational Lens Efficient Explorer \citep[\glee,][]{Suyu:Halkola:2010} in interactive mode. 
We further test the robustness of our lens mass model by comparing the directly measured stellar velocity dispersions ($\sigma_{\rm obs}$) of deflectors with their model-predicted velocity dispersions ($\sigma_{\rm pred}$).

Section \ref{Sec:Data} provides more details about the \agel \ survey, the lens systems studied here, the images used for lens modeling, and the stellar velocity dispersion measurements.  
Section \ref{Sec:modelling} describes the lens modeling process using GLEE.
Section \ref{Sec:Results} presents our lens modeling results, 
comparison of $\sigma_{\rm pred}$ of deflector galaxies with $\sigma_{\rm obs}$, and the updated $\gamma$--$z$ diagram.
Section \ref{sec: Discussion} discusses the evolution of $\gamma$ with $z$ obtained by simulations, purely dynamical observations, and lensing observations, as well as  possible reasons behind discrepancy among these studies. 
In this section, we also compare the lensing-only density profile slopes with the slopes obtained via L\&D analysis for our sample.
Finally, Section \ref{Sec:conclusions} presents the conclusion of this work and further plans.

Cosmological constants for estimating distances and scales used in this paper are $\Lambda \rm CDM$ $H_{0}=70 \,\rm km \, s^{-1} \, Mpc^{-1}$, $\Omega_{\rm m}=0.3$, and $\Omega_{\rm vac}=0.7$.

\section{Data} 
\label{Sec:Data} 

\begin{figure*}
\begin{center}
\includegraphics[clip=true,trim= 30mm 10mm 03mm 05mm,width=  1.1\textwidth]{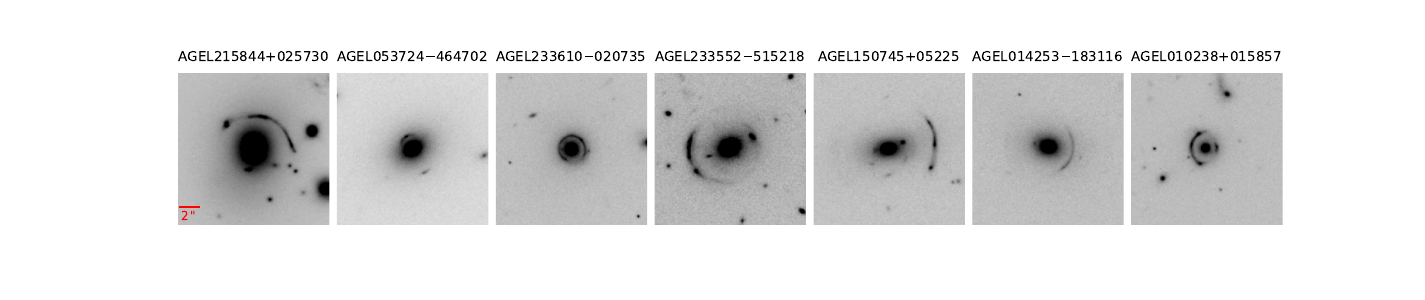}
\caption{HST/WFC3 images of the seven \agel \, lenses modeled here in F140W filter. All seven images are $16\arcsec \times 16\arcsec$ in size with a pixel resolution of $0.08 \arcsec/\rm pixel$. The red scale bar indicates two arcsecond scale that is same for all images. Section \ref{Sec:Data} provides more details about the \agel \ naming of these lens systems,  deflector and background source redshifts, HST image observations programs,  and the kinematic observations needed to test our lens modeling. }
\label{fig:HST images}
\end{center}
\end{figure*}

The complete \agel \, survey comprises about 1900 lens candidates \citepalias[see section 2.1.2 in][]{Tran:Harshan:2022} selected using convolutional neural networks from Dark Energy Survey (\des) and Dark Energy Camera Legacy Survey (\decals). 
\agel \, survey has spectroscopically confirmed about 100 strong lenses which include galaxy-scale, galaxy-group, and cluster-scale lens systems, with deflectors at $0 \lesssim z \lesssim 1$. Of the 100 confirmed lenses, redshifts for 53 are already published in \citetalias{Tran:Harshan:2022}, and the remaining lenses will be presented in \agel \, Data Release 2 (DR2, in preparation). 

To study galaxy evolution through the observed evolution of total mass profile with redshift (cosmic time), 
one needs galaxy-scale strong gravitational lenses where deflectors are single galaxies. 
About $50\%$ of confirmed \agel\ lens systems are galaxy-scale lenses, $50\%$ of which already have high-resolution Hubble Space Telescope (HST) images. Of the confirmed galaxy-scale lenses with HST imaging, we have 15 lens systems with deflectors at spectroscopic redshift $ \gtrsim 0.5$ and an additional 14 lenses with photometric redshift $\gtrsim 0.5$. Follow-up observations are underway to obtain spectroscopic redshifts, HST images, and spectra for velocity dispersion measurements for more lenses.

Depending on the availability of 1) high-resolution HST images 2) deflector and source spectroscopic redshifts and 3) spectra for deflector's  stellar velocity dispersion measurements or velocity dispersion from literature, we selected an initial sample of seven galaxy-scale lenses for this work. 
The first two are required for lens modeling and the velocity dispersion is required to check the robustness of our lens modeling results.
The galaxy-scale \agel \ lenses studied in this paper are \agel215844$+$025730 (hereafter \agel2158), 
\agel053724$-$464702 (hereafter \agel0537),
\agel233610$-$020735 (hereafter \agel2336), 
 \agel233552$-$515218 (hereafter \agel2335), \agel150745$+$052256 (hereafter \agel1507),
 \agel014253$-$183116 (hereafter \agel0142), and \agel010238$+$015857 (hereafter \agel0102).
Here, the numbers after ``\agel" comprise deflector galaxy coordinates, RA in hour:minute:second (hhmmss) and DEC in degree:minute:second (+/-ddmmss).

For lens modeling, we used the HST images taken by the Wide-Field Camera 3 (WFC3) in filter F140W from the SNAP programs $\#16773$ (Cycle 29, P.I. K Glazebrook) and $\#15867$ (Cycle 27, P.I. X Huang). 
The images are available at Mikulski Archive for Space Telescopes (MAST) at the Space Telescope Science Institute (STScI) and can be accessed via \dataset[10.17909/z66n-v326]{https://doi.org/10.17909/z66n-v326}.
Program $\#16773$ observed the lens systems in the F140W filter for three exposures of 200 s (and also in the UVIS/F200LP filter for two exposures of 300 s). 
This multi-filter observing sequence within one truncated HST orbit was optimized using \textsc{LensingETC} \citep{Shajib:Glazebrook:Barone:2022}. 
Program $\#15867$ observed the lens systems for three exposures of 399.23 s  in the F140W filter.
The data was reduced using the \textsc{AstroDrizzle} software package from STScI \citep{Avila:Hack:2015}. The drizzled pixel size in the F140W filter was set to $0.08\arcsec$.
The HST images of seven \agel \, lenses modeled in this paper are presented in Figure \ref{fig:HST images}.
%

The deflector redshift (\zdefl), source redshift (\zsrc), kpc-to-arcsecond scale at the deflector plane and the line-of-sight stellar velocity dispersion ($\sigma_{\rm obs}$) of the deflector galaxy in our lens sample are provided in Table \ref{table:redshift}.
Deflector redshifts for \agel2158, \agel2336, \agel1507, and \agel0102
are taken from the Baryon Oscillation Spectroscopic Survey (\boss) of Sloan Digital Sky Survey \citep[\sdss-\boss,][]{Eisenstein:Weinberg:2011}.
Source and deflector redshifts for remaining lenses, except for AGEL0537, are taken from \citetalias{Tran:Harshan:2022}. 
Redshifts for AGEL0537 are measured by us using spectra obtained by X-shooter spectrograph \citep{Xshooter:Vernet:2011} at the ESO Very Large Telescope (VLT).

The single aperture line-of-sight stellar velocity dispersions, $\sigma_{\rm obs}$, for the deflector galaxies of \agel2158, \agel2336, and \agel1507 are taken from \sdss-\boss \, measurements  available in the literature \citep{Thomas:Steele:2013}. 
For the remaining lenses, we measured deflector's $\sigma_{\rm obs}$ from their absorption line spectra with the help of the Penalized Pixel-Fitting stellar kinematics extraction \citep[\textsc{pPXF},][]{Cappellari:ppxf:2012, Cappellari:2023} software. 
For \agel0537, \agel2335, and \agel0142, deflector spectra were obtained using VLT/X-shooter.
Spectra for \agel0102 were obtained using the Echellette Spectrograph and Imager \citep[ESI,][]{ESI:Sheinis:2002} at the Keck II telescope.
To measure the velocity dispersions using \textsc{pPXF}, we used templates from the 
Medium-resolution Isaac Newton Telescope library of empirical spectra \citep[MILES,][]{MILES:Sanchez-Blazquez:2006} and a 12th order additive Legendre polynomial to correct low-frequency continuum variations and remove residuals resulting from minor flux calibration errors \citep[following][]{vandeSande:Bland-Hawthorn:2017}. 
The seeing conditions for these observations, which are later required to predict velocity dispersions based on our lens modeling results, are listed in Table \ref{table:redshift}.

\rotate
\begin{deluxetable*}{lrrrrrrrlrrrrrr} 
\tabletypesize{\small}
\setlength{\tabcolsep}{2pt}
 \tablehead{
 Name & \zdefl & \zsrc & $\rm scale_{defl}$ & \rm $\rm size_{im}$ &  $\rm size_{src}$  & $\sigma_{\rm obs}$ & Aperture & Instrument & Seeing & $\sigma_{\rm pred}$ & $\rm \mathfrak{m}_{\rm ap,gals}$  & $R_{\rm half,gal}$ & $\gamma^{\rm lens}$ & $\gamma^{\rm L\&D}$\\ 
 &  &  &  $\rm kpc/\arcsec$ & $\arcsec$ & $\arcsec$ & km/s & $\arcsec$ & & $\arcsec$ & km/s & mags & $\arcsec$ &  & \\  
 (1) & (2) & (3) & (4) & (5) & (6) & (7) & (8) & (9) & (10) & (11) & (12) & (13) & (14) & (15) }
\startdata
 \agel2158 & 0.28654\tablenotemark{a} & 2.08015 & 4.319 & 11.2 & 1.83$\times$2.04 & 322$\pm$17 & 2 & SDSS-BOSS & 1.4 & 315$\pm$19 & 16.11$\pm$0.02 & 2.428$\pm$0.087 & 1.820$\pm$0.014 & 1.91$\pm$0.03 \\ 
 \agel0537 & 0.35200 & 2.34430 & 4.940 & 7.04 & 0.63$\times$0.72 & 288$\pm$19 & 11$\times$1.2 & Xshooter  & 0.71 & 237$\pm$7 & 17.17$\pm$0.03 &1.007$\pm$0.050 & 1.810$\pm$0.016 & 2.10$\pm$0.05\\
 \agel2336 & 0.49417\tablenotemark{a}  & 2.66173 & 6.065 & 4.96 & 0.84$\times$1.00 & 272$\pm$35 & 2 & SDSS-BOSS & 1.4 & 285$\pm$10 & 17.97$\pm$0.04 & 0.687$\pm$0.043 & 2.248$\pm$0.042 & 2.20$\pm$0.10\\ 
 \agel2335 & 0.56600 & 2.22450 & 6.502& 8.96 & 1.54$\times$2.09 & 321$\pm$27 & 11$\times$1.2 & Xshooter & 0.85 & 426$\pm$15 & 17.37$\pm$0.08 & 1.984$\pm$0.301 & 2.020$\pm$0.018 & 1.78$\pm$0.05\\ 
 \agel1507 & 0.59454\tablenotemark{a}  & 2.16275 & 6.654 & 11.2 & 1.41$\times$1.16 & 303$\pm$38 & 2 & SDSS-BOSS & 1.4  & 322$\pm$27 & 17.70$\pm$0.03& 1.432$\pm$0.060 & 1.620$\pm$0.018 & 1.66$\pm$0.07\\ 
 \agel0142 & 0.63627 &  2.46972 & 6.868& 8.00 & 1.46$\times$1.60 & 316$\pm$40 & 11$\times$1.2 & Xshooter & 0.70 & 385$\pm$29 & 17.59$\pm$0.06 & 1.460$\pm$0.165 & 2.208$\pm$0.100 & 1.97$\pm$0.10\\
 \agel0102 & 0.86690\tablenotemark{a}  & 1.81696 & 7.708 & 4.96 & 0.94$\times$1.01 & 291$\pm$51 & 20$\times$1 & ESI & 0.8 & 339$\pm$12 & 17.30$\pm$0.05 & 0.981$\pm$0.083 & 1.944$\pm$0.028 & 1.72$\pm$0.20\\
\enddata
\caption{\label{table:redshift} \agel \ pilot lens sample. 
Columns: (1) system name, (2) and (3) deflector and source spectroscopic redshifts, respectively, all with a typical uncertainty of $\lesssim 0.00005$,  (4) kpc-to-arcsec scale for the deflector plane obtained using the cosmological parameters assumed in Section \ref{Sec:Introduction}, (5) image square grid cutout size, (6) source rectangular (length$\times$width) grid cutout size, (7) directly measured stellar velocity dispersion of deflector galaxy, (8) and (9) aperture size and the instrument used for $\sigma_{\rm obs}$; SDSS-BOSS with shell aperture of diameter $2\arcsec$, Xshooter and ESI with rectangular slit apertures (10) seeing disk full width at half maximum size for the spectroscopic observations (used to predict lensing velocity dispersions), (11) stellar velocity dispersion of deflector predicted using \textsc{lenstronomy}'s \textsc{GalKin} routine based upon our most probable lens models and assumed anisotropy (see Section \ref{subsec: predicting velcocity dispersion}), (12) and (13) apparent magnitude of the deflector galaxy in ABMAG system and overall deflector galaxy half-light radius along the major axis, respectively, obtained by integrating the double S\'ersic fits obtained during lens light modeling, (14) magnitude of slope of power-law mass density profile ($\rho \propto r^{-\gamma^{\rm lens}}$) for deflector galaxies obtained using lensing-only analysis (15) magnitude of the slope of the power-law mass density profile obtained using joint lensing and dynamical analysis assuming isotropic stellar orbits.}
\tablenotetext{a}{Taken from \boss \,\citep{Eisenstein:Weinberg:2011}.}
\end{deluxetable*}

\section{Lens Modeling} 
\label{Sec:modelling}
Gravitational lensing is the deflection of light coming from a background source by a foreground mass distribution that forms magnified and distorted images of the background source. In case of strong lensing multiple images of the background source are observed. 
The foreground mass is usually referred to as the deflector or the lens.
During lens modeling, we constrain the mass density profile of the deflector via the lens equation: $\Vec{\beta} = \Vec{\theta} - \alpha (\Vec{\theta})$, and reconstruct the background source's intrinsic image.
Lens equation links the lens plane (\textit{aka} observed image plane) coordinates, $\Vec{\theta}$, with that of background source plane, $\Vec{\beta}$, via the scaled deflection angle $\alpha (\Vec{\theta})$ that is associated with the gradient of the lens potential $\alpha (\Vec{\theta})=\nabla \psi(\Vec{\theta})$.

Essentially we measure the dimensionless surface (projected) lens mass density, also known as  convergence, denoted by $\kappa$.
Convergence is related to the lens potential via  Poisson equation $2\kappa = \nabla^2 \psi(\Vec{\theta})$.
Generally, convergence is expressed in the units of critical density, $\Sigma_{\rm crit} = \frac{c^2 D_{\rm s}}{4 \pi G D_{\rm d} D_{\rm ds}}$, of a lens system. 
Here, $D_{\rm s}$, $D_{\rm d}$, and $D_{\rm ds}$ are  angular diameter distances between observer and source, observer and deflector, and deflector and source, respectively.
The scaled deflection angle is related to the actual deflection angle, $\hat{\alpha} (\Vec{\theta})$, via 
$\alpha (\Vec{\theta})=D_{\rm ds}/D_{\rm s} \hat{\alpha} (\Vec{\theta})$. See \citet{Schneider:Ehlers:Falco:1992} for a detailed derivation of lensing equations from the General Relativity.

We used Gravitational Lens Efficient Explorer \citep[\glee,][]{Suyu:Halkola:2010, Suyu:Hensel:McKean:2012} for lens modeling, which involves modeling the deflector total mass profile, deflector light profile, and reconstructing the background source surface brightness.
GLEE is an interactive modeling software which uses Bayesian optimization algorithms like simulated annealing and Markov chain Monte Carlo (MCMC) methods for parameter estimation \citep{Kirkpatrick:Gelatt:1983, Dunkley:Bucher:2005}.
\glee \ is used for ``Lensing-only" lens modeling where we only need a high-resolution image to perform modeling in single-plane mode. Later, we require source and deflector redshifts to obtain the deflector mass profile from the unitless convergence profile.
We also use additional kinematic information, such as the observed stellar velocity dispersion of the deflector galaxy, to check the accuracy of the deflector mass model. 
In following sections, we describes the mass and light model parameterizations that we used, input data/files required, and the  process of lens modeling with \glee.

\subsection{Models parameters}
\label{subsec:model para}
To model the deflector's projected total mass density profile, we used Softened Power-Law Elliptical Mass Distribution \citep[SPEMD,][]{Barkana:1998} along with an external shear component. 
In general, SPEMD profiles have a constant density within a core radius $R_{\rm c}$, followed by an elliptically symmetric power-law fall off.
Parameters for the SPEMD profile based on the \textsc{fastell} code from \citet{Barkana:1998} are: 
centroid position in arcsec ($x_{\rm m}, y_{\rm m}$), mass axis ratio ($q_{\rm m}$), 
mass position angle ($\varphi_{\rm m}$ in radians, measured counter-clockwise from positive x-axis), projected Einstein radius ($R_{\rm Eins}$  in arcsec) along the major axis, the radial power-law slope  ($-\gamma^{\scaleto{\rm lens}{4pt}}$), and the core radius $R_{\rm c}$ set to $10^{-4}$ arcsec in this work.

Convergence, $\kappa$, for the SPEMD profile for a very small core radius, under the $\lim_{R_{\rm c}\to 0} \kappa (x,y)$, is expressed as 
\begin{equation}
\kappa (x,y) = \frac{(3-\gamma^{\scaleto{\rm lens}{4pt}})}{1+q_{\rm m}} \left [ \frac{R_{\rm Eins}}{\sqrt{(x-x_m)^2 + \frac{(y-y_m)^2}{q_{\rm m}^2}}} \right ]^{\gamma^{\scaleto{\rm lens}{4pt}}-1}.
\label{eqn: SPEMD convergence}
\end{equation}
The mass distribution is rotated by the mass position angle $\varphi_{\rm m}$ during modeling.
Here the parameter value $\gamma^{\scaleto{\rm lens}{4pt}} = 2$ corresponds to an isothermal three-dimensional mass density profile ($\rho \propto r^{-2}$). 
The spherical equivalent Einstein radius, $\theta_{\rm Eins}$, is related to $R_{\rm Eins}$ via $\theta_{\rm Eins}= (2/(1+q_{\rm m}))^{1/(\gamma^{\scaleto{\rm lens}{4pt}} -1)} \sqrt q_{\rm m} R_{\rm Eins} $  \citep[see][]{Suyu:Auger:Hilbert:2013}.

For point sources, the magnification ($\mu$) due to lensing is dependent on the convergence and shear ($\gamma_{\rm tot} (\gamma_1, \gamma_2)= \sqrt{\gamma_{1}^2 + \gamma_{2}^2}$) caused by the deflector. It is given by the determinant of magnification tensor ($\rm det \, (A^{-1})$) as 
$\mu = \rm det \, (A^{-1}) = 1/((1-\kappa)^2 - \gamma_{1}^2 - \gamma_{2}^2)$. 
However, the magnification of an extended elliptical or irregular background source depends on the magnification matrix ($\kappa, \gamma_{\rm tot}$) as well as the intrinsic surface brightness distribution of the background source  which is not already known \citep{Treu:2010, Birrer:2021}.

The additional external shear accounting for the tidal gravitational potential of deflector's local environment is parameterized using external shear magnitude ($\gamma_{\rm ext}$) and position angle ($\varphi_{\rm ext}$) measured anti-clockwise from the positive $x$-axis. 
The lens potential associated with the external shear \citep[taken from][]{Suyu:Halkola:2010} can be expressed as 
\begin{equation}
\psi_{\rm ext}(x,y)=\frac{1}{2} \gamma_{\rm ext} \{(x^2 +y^2) \cos{(2\varphi_{\rm ext})} + 2x y \sin{(2 \varphi_{\rm ext})} \}.
\label{eqn: external shear}
\end{equation}
Where $\varphi_{\rm ext}=0$ means the lensed image is stretched horizontally, and $\varphi_{\rm ext}=\pi/2$ means the image is stretched vertically.

To model the deflector light profile, we use the S\'ersic function \citep{Sersic:1963} which defines the apparent intensity at any point ($x,y$) at an elliptical isophote at radius $R(x,y)=\sqrt{(x-x_{\scaleto{\rm L}{4pt}})^2 + (y-y_{\scaleto{\rm L}{4pt}}/q_{\scaleto{\rm L}{4pt}})^2}$ from the light/photometric center ($x_{\rm L}, y_{\rm L}$) of a galaxy as,
\begin{equation}
I(x,y)=A \exp \left\{-b_{n} \left[\left(\frac{R(x,y)}{R_{\rm eff}} \right)^{1/n} -1 \right] \right\}.
\label{eqn: Sersic}
\end{equation}
Here, $q_{\scaleto{L}{4pt}}$ is the galaxy's photometric minor versus major axis-ratio, $n$ is the  S\'ersic index representing profile shape,  $A$ is called profile amplitude representing intensity at  $R_{\rm eff}$, and $b_{\rm n}$ is a S\'ersic index-dependent normalization constant such that $R_{\rm eff}$ represents the half-light radius in the direction of the semi-major axis. 
The value of $b_n$ is calculated by solving $\Gamma(2n)=2\gamma(2n, b_{n})$, where $\Gamma(2n)$ is the complete gamma function and $\gamma(2n, b_{n})$ is the  incomplete gamma function. 
It can also be approximated by $b_{\rm n} \approx 2n- (1/3) + (4/405n) + (46/25515n^2)$ for $0.36 < n < 10$ \citep{Ciotti:Bertin:1999, Dutton:Brewer:2011}.

\subsection{Input files for lens modeling}
\label{subsec: input files}
Lens modeling with \glee \, requires many input files such as the lens image cutout, point spread function (PSF) for the image, arc mask, lens/deflector mask, and the error map for the input lens image. 
Image cutout sizes ($ \rm size_{\rm im}$) for our galaxy-scale lenses are mentioned in Table \ref{table:redshift}, these are about 4-5 times the galaxy half-light radius ($R_{\rm half, gal}$). 
The cutout sizes are selected to include lensed sources (Einstein radius $\lesssim 3-4\arcsec$) and the immediate environment of the deflectors ($\lesssim 4\arcsec \, \rm to \, 6\arcsec $).
We obtained PSF for the HST images using \textsc{tinytim} \citep{Krist:2011}. 

The arc mask is a FITS file marking the pixels containing lensed source e.g., arc plus counter-image, in the lens image cutout. 
The lens mask is a FITS file masking pixels containing the surrounding luminous objects, e.g., foreground stars, which are not a part of the deflector potential.
We generated the masks manually by marking the mask regions using DS9 followed by region file to FITS file conversion. 
Arc and lens masks are important to avoid light contamination from the lensed sources, foreground stars, and nearby bright galaxies when modeling the lens (deflector) surface brightness. 
Arc mask is further used to mark lensed background source regions required for source reconstruction.

We obtained the error maps by adding the background noise ($\sigma_{\rm bkg}$) and the Poisson noise ($\sigma_{\rm P}$) in quadrature for each pixel ($\sqrt{\sigma_{\rm bkg}^2 + \sigma_{\rm P}^2}$). 
Here, the background noise is the standard deviation calculated from an empty region of sky in the science image, and the Poisson noise for each pixel is the reduced image intensity (in counts per second) over exposure time ($ \sigma_{\rm P}=\sqrt{\rm intensity/exposure\,time}$).
A detailed description of these input files and their preparation can also be found in \citet[][section 2.1.1]{Ertl:Schuldt:2023}. 

\subsection{Modeling with \glee }
\label{subsec: modeling method}
Lens modeling using \glee\ is performed in three phases. 
First, using the ``Position modeling," we obtain an initial guess for the lens mass (SPEMD) profile parameters.
Second, we perform the lens light modeling which is independent of mass model. 
Third, we perform ``Extended source modeling," which involves lens mass modeling and source reconstruction.
These three phases are described in detail below. 

We assume uniform priors on the lens mass and light profile parameters.
The parameters obtained from the first phase, position modeling, enable faster convergence than when using extended source modeling directly. 
However, lens mass profile parameter likelihood from the position modeling phase is not included in the final likelihood during extended image modeling which provides the final most probable lens mass model.

We use simulated annealing for parameter optimization during position modeling.  
However, for efficient sampling/optimization during lens light modeling and extended source modeling, we use the \textsc{emcee} ensemble sampler \citep{Foreman-Mackey:Hogg:2013} and the Metropolis-Hastings (M-H) MCMC algorithm.
The \textsc{emcee} routine which is highly parallelizable is used to obtain the first samples and first sampling covariance matrix. 
Subsequently, use of the M-H MCMC algorithm with the covariance matrix enables faster convergence, where convergence is tested by the power spectrum method from \citet{Dunkley:Bucher:2005}.

\subsubsection{Position modeling with \glee}
\label{subsubsec:position}
Position modeling constrains the lens mass distribution parameters using the lensed image and source positions with respect to the deflector position \citep{Halkola:2006, Halkola:2008}. 
We identify the positions of the multiple images of the background source (i.e., the centroid/photo peak) of the lensed galaxy or a bright star-forming clump) using DS9 visualization.
Position modeling has two steps.

First, the lens equation ($\Vec{\beta}  = \Vec{\theta} - \alpha (\Vec{\theta}) $) is used to predict the intrinsic source position ($\Vec{\beta}$) using the observed multiple positions of the lensed  source ($\Vec{\theta}$) and the scaled deflection angle ($\alpha (\Vec{\theta})$) based upon the prior deflector mass profile parameters.
This maps the lensed image positions back to the source plane via the lens equation. 
Each lensed source position predicts an intrinsic source position, and the final intrinsic source position is taken as their magnification-weighted average.
Further, the lens mass profile parameters (linked with the deflection angle as described in section \ref{subsec:model para}) are varied to minimize the source position $\chi^2_{\rm pos}$ defined in \citet[][their equation 7]{Halkola:2006}.

The second step optimizes deflector mass profile parameters using the observed image peak positions as constraints. 
Here, the lensed source positions are predicted based on the intrinsic source position and lens mass model  obtained in the previous step. 
Further, the lens mass model parameters are varied to minimize image position $\chi^2_{\rm pos}$ \citep[][their equation 6]{Halkola:2006} based on the difference between the observed and predicted lensed source positions.  

\subsubsection{Lens light modeling with \glee}
\label{subsubsec:lens light}
Lens light modeling captures the deflector's surface brightness distribution by fitting one or more S\'ersic functions directly to the deflector galaxy image (Eq. \ref{eqn: Sersic}). 
This phase uses the arc and lens masks to block lensed source light and surrounding contaminants in the lens image cutout and captures only the deflector light.

We found that, two S\'ersic functions efficiently fit the deflector light in our galaxy scale lenses, leaving behind a uniform residual.
The most probable lens light model parameters are obtained by minimizing the lens surface brightness $\chi^2_{\rm SB}$ provided in \citet[][their equation 5]{Ertl:Schuldt:2023}.
The lens light model obtained here is further used to separate the lens light overlapping with the arc in the next phase so that the arc light without contamination from the deflector can be used for source reconstruction. 
For the same reason, we mask out any other bright object near or overlapping with the arc. 

\subsubsection{Extended source modeling with \glee}
\label{subsubsec:extended}
This phase involves constraining the lens mass profile and external shear by reconstructing the surface brightness profile of the background source \citep[following][]{Suyu:Marshall:Hobson:2006, Suyu:Halkola:2010}. 
Here, we improve upon the lens mass model obtained from the position modeling with the help of multiply lensed extended background source surface brightness distributions which offer tighter constraints on the mass model.

\glee\ performs a pixellated source reconstruction on a grid of fixed pixel dimensions following conservation of surface brightness. 
The intrinsic source image is reconstructed via Bayesian regularized linear inversion of lensed source surface brightness. 
The regularization functions can be curvature or gradient depending on whether the lensed source is intrinsically smooth or clumpy.
For all the lenses presented here, as the lensed sources appear to be featureless and smooth, we used  the ``curvature" regularization for reconstructing the source surface brightness distribution. 

The source grid size ($ \rm size_{\rm src}$), mentioned in Table \ref{table:redshift}, is the minimal rectangular region on the source plane that contains the area to which the arc mask on the image plane maps (via the lens equation).
The dimensions of the source grid are chosen such that the source pixel resolution is approximately the image pixel resolution divided by the square root of the average magnification \citep[see][]{Suyu:Marshall:Hobson:2006}.

The expression for the likelihood of the lens mass model parameters, which is equivalent to the Bayesian evidence of source reconstruction, is provided in \citet[][their equation 12]{Suyu:Halkola:2010}. 
The best-fit or the most probable lens model is selected when the reduced chi-square ($\chi^2_{\rm red}$) for extended source reconstruction lies between  0.95 and 1.05. 
$\chi^2_{\rm red}$ is dependent on the Bayesian evidence of the source reconstruction, Likelihood of the lens light fitting and the effective degree of freedom, it is calculated in \glee \ following \citet{Suyu:Marshall:Hobson:2006}.

\subsection{Model-predicted stellar velocity dispersion}
\label{subsec: predicting velcocity dispersion}
Lens modeling based solely on the imaging observables can have various degeneracies related to model parametrizations or intrinsic to data \citep[see][]{Shajib:Vernardos:Collett:2022}.
Importantly, the lens mass model is affected by the Mass-Sheet Degeneracy (MSD).
MSD is a result of a multiplicative transformation of the lens equation 
known as the mass-sheet transformation \citep[MST,][]{Falco:Gorenstein:1985}, which can alter the mass distribution (e.g., $\kappa$ map) without affecting the resultant lensed image configuration \citep{Schneider:Sluse:2013}. 
MST also suggests the possibility of non-power law mass models  \citep[see][]{Birrer:Shajib:Galan:2020}.

Only the prior knowledge of source intrinsic size or intrinsic magnification or an independent measure of lensing potential can break this degeneracy.
Therefore, it is vital to check the modeling results using additional kinematic observations which can provide an independent measure of the lensing potential.
One way to check the lens modeling results is by comparing the lens model predicted velocity dispersion, $\sigma_{\rm pred}$, with the observed velocity dispersion, $\sigma_{\rm obs}$.

 
We used the \textsc{GalKin} module of  \textsc{lenstronomy} \citep{Birrer:Amara:2018, Birrer:Shajib:Gilman:2021} software to obtain $\sigma_{\rm pred}$.
\textsc{GalKin} constrains $\sigma_{\rm pred}$ via  
spherical Jeans modeling with the help of deflector's mass model, light profile, and stellar anisotropy for a given observing condition.  
While calculating $\sigma_{\rm pred}$ for a deflector galaxy, we used the same aperture size, aperture type, and the seeing disk Full Width at Half Maximum (FWHM) size as that in the observing condition for $\sigma_{\rm obs}$.
The aperture type (circular shell or rectangular slit), aperture size centered around the galaxy centroid, and the seeing FWHM for $\sigma_{\rm obs}$ of deflectors in our lens sample are presented in Table \ref{table:redshift}.

The anisotropy distribution of stellar orbits is degenerate with the mass distribution and is often referred as the mass-anisotropy degeneracy \citep{Binney:Mamon:1982}. 
Thus, one has to assume an anisotropy distribution to predict the velocity dispersion.
For our sample, we use the Osipkov-Merritt \citep{Osipkov:1979, Merritt:1985} radial anisotropy distribution, parameterized as $\beta_{\rm ani}=r^{2}/(r^{2}_{\rm ani} +r^{2})$. 
Here, $r_{\rm ani}$ is the free parameter determining the degree of anisotropy.
The choice of anisotropy is critical to the predicted stellar velocity dispersion because a higher anisotropy can result in an overestimated stellar velocity dispersion and vice-versa \citep[see Fig B.1 in][]{Birrer:Shajib:Galan:2020}. 

Many observations suggest that massive elliptical galaxies are generally isotropic to mildly radially anisotropic at larger radii, suggesting $\beta_{\rm ani} \lesssim 0.3$  \citep{Binney:1978, Kronawitter:Saglia:2000, Saglia:Kronawitter:2000, Chen:Hwang:2016}.
Relatively oblate rotating ellipticals or lenticular galaxies can be isotropic in their central regions; however, they become radially anisotropic beyond a certain radius such that $\beta_{\rm ani} \lesssim 0.5$ \citep{Cappellari:Emsellem:2007}. 
We vary the anisotropy radius uniformly between 1 to 10 times the galaxy's spatial (3D) half-light radius $r_{\rm eff}$, which corresponds to $0.01 \lesssim \beta_{\rm ani}(r_{\rm eff}) \lesssim 0.5$, consistent with \citet{Koopmans:Bolton:2009} based upon 58 ETGs in the  Sloan Lens ACS Survey sample \citep[SLACS,][]{Bolton:Burles:2006}.

\section{Results}
\label{Sec:Results}

\subsection{Lens models obtained with \glee}
\label{Result:lens_models}
\begin{figure}
\begin{center}
\includegraphics[clip=true,trim= 01mm 03mm 00mm 00mm,width=  0.5\textwidth]{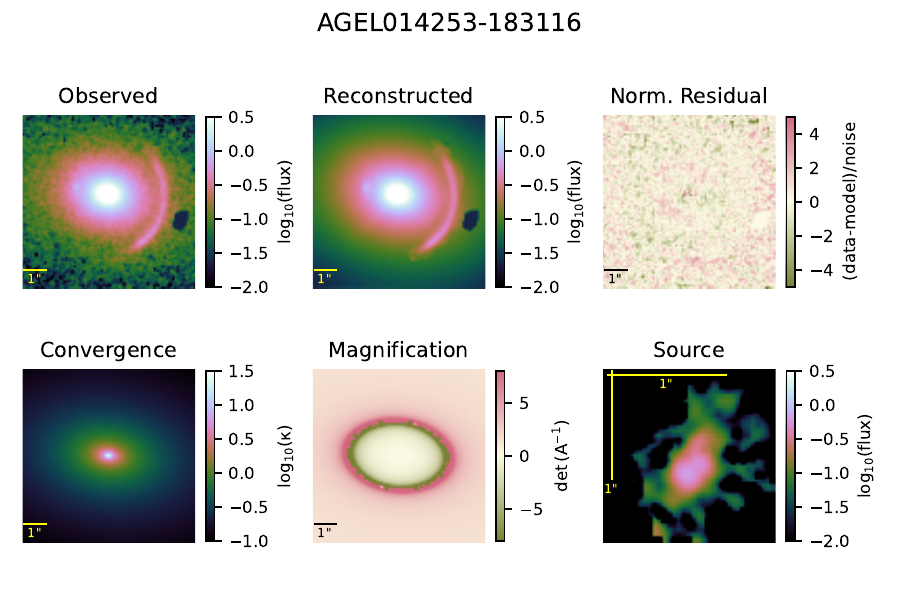}
\caption{Lens modeling results for AGEL0142 obtained using \glee. Panels from top left to right represent the observed lens configuration, reconstructed lens model, and normalized residual. Panels from bottom left to right are convergence map of the deflector's density profile, magnification model, and the reconstructed background source.
Here, flux is in the units of electron counts per second, and convergence and magnification maps are unitless (described in section \ref{Sec:modelling}).  The black patches on the observed and reconstructed panels are pixels that are masked out during modeling.
The yellow and black lines provide one arcsec scale in the deflector and source planes. The image and source grid angular sizes are listed in Table \ref{table:redshift}.}
\label{fig:models}
\end{center}
\end{figure}

Aiming to measure the total mass density profile for the deflectors in our lens sample, we modeled all seven \agel \, lenses following the ``Lensing-only" method described in Section \ref{Sec:modelling} such that $\chi^2_{\rm red}$ for the final best-fit model is between $0.95-1.05$.
As depicted in Figure \ref{fig:models} for \agel0142 and in Figure \ref{fig:models appendix} for the remaining six lenses, we successfully reconstructed the observed lensing configuration and  the background source surface brightness for all seven systems.
In Figures \ref{fig:models} and \ref{fig:models appendix}, we show the  observed HST image, the most probable lens model, normalized residual\footnote{The normalized residual for each pixel was obtained by dividing the difference between the observed image and the lens model by estimated standard deviation.}, convergence map ($\kappa$) for the deflector SPEMD density profile, magnification model ($\mu$), and the background source reconstruction.
The first five panels in Figure \ref{fig:models} have the same grid size and pixel resolution ($0.08 \rm \ arcsec/pixel $). 
The source grid in the bottom right-most panel has a higher resolution ($0.04 \rm \ arcsec/pixel $) dependent on the observed image resolution and magnification caused by the lens \citep{Suyu:Marshall:Hobson:2006}.

Parameters for deflector's convergence profile (Eq. \ref{eqn: SPEMD convergence}), external shear (Eq. \ref{eqn: external shear}), and deflector light profile (Eq. \ref{eqn: Sersic}) are provided in Table \ref{table:lens parameters}.
The subscripts 1 and 2 in Table \ref{table:lens parameters} refer to the two S\'ersic components of deflector light profile. 
All the parameter estimates are the median values of their one-dimensional posterior probability density function from the final  MCMC chain. 
The uncertainties presented here are based on the 16th and 84th percentiles of the distribution, representing $68\%$ ($1 \sigma$) bound around the median value.

We additionally measured the apparent magnitude ($\mathfrak{m}_{\rm ap, gal}$, in the HST F140W filter) and the projected half-light radius along the major axis ($R_{\rm half,gal}$) for the deflector galaxy in all seven lens systems  (see Table \ref{table:redshift}). 
We obtained the total galaxy flux by integrating the two S\'ersic components of the deflector's light profile over the image plane.
Further, the flux in the units of electron count per second ($e^-/s$) is converted to AB system magnitude following HST \href{https://hst-docs.stsci.edu/wfc3dhb/chapter-9-wfc3-data-analysis/9-1-photometry}{WFC3 Data Analysis}.
We numerically calculated the effective half-light radius along the major-axis, $R_{\rm half,gal}$, of the deflector galaxy by integrating the two S\'ersic components and looking for the (isophote) semi-major axis radius enclosing half of the total galaxy flux.  
The equivalent circularized half-light radius, $R_{\rm half, gal, eq}$, is also presented in Table \ref{table:lens parameters}.

\subsection{Comparing model-predicted and observed velocity dispersions}
\label{Result: velocity dispersion}
\begin{figure}
    \centering
    \includegraphics[clip=true,trim= 03mm 2.5mm 01mm 2.5mm,width=  0.48\textwidth]{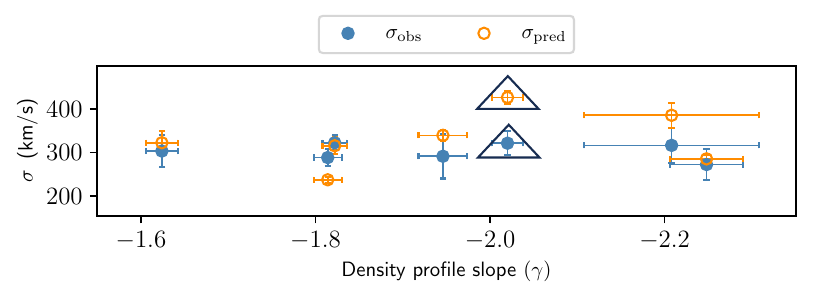}
    \caption{Observed (blue filled circles) and model-predicted  (orange open circles) stellar velocity dispersions of deflector galaxies in the  lens systems modeled in this paper, plotted against their total mass density profile slope ($\gamma= -\gamma^{\scaleto{\rm lens}{4pt}}$). 
    For all lenses, except for AGEL2335 marked by black triangles, $\sigma_{\rm pred}$ is consistent with $\sigma_{\rm obs}$ (see Section \ref{Result: velocity dispersion}). Values of $\sigma_{\rm obs}$, $\sigma_{\rm pred}$, and $\gamma^{\scaleto{\rm lens}{4pt}}$ 
    are provided in Table \ref{table:redshift}.}
    \label{fig:sigma-gamma}
\end{figure}
 
To check the accuracy of the deflector/lens mass density profile obtained via lens modeling, it is important to compare the model predicted stellar velocity dispersion, $\sigma_{\rm pred}$, with the observed velocity dispersion  $\sigma_{\rm obs}$.
A significant difference between $\sigma_{\rm pred}$ and  $\sigma_{\rm obs}$ may be an indication of an inappropriate lens mass model due to parameter degeneracy (see Section \ref{subsec: predicting velcocity dispersion}), a potential line-of-sight mass overdensity (under density) if $\sigma_{\rm pred}$ is higher (lower) than $\sigma_{\rm obs}$ \citep{Li:Wang:2018}, or  a possible deviation from the power-law mass model \citep{Birrer:Shajib:Galan:2020}. 
On the other hand, the consistency between $\sigma_{\rm obs}$ and $\sigma_{\rm pred}$ suggests that the lens models very well represent the actual mass distribution of deflector galaxies.

Following Section \ref{subsec: predicting velcocity dispersion}, we calculated $\sigma_{\rm pred}$ for deflector galaxies based on their most probable lens model parameters presented in Table \ref{table:lens parameters}. 
A plot comparing $\sigma_{\rm obs}$ and $\sigma_{\rm pred}$ against the slope of the total density profile of deflector galaxies is shown in Figure \ref{fig:sigma-gamma}.
Both $\sigma_{\rm obs}$ and $\sigma_{\rm pred}$ values for all lenses are provided in Table \ref{table:redshift}. 
For \agel2158, \agel2336, \agel1507, \agel0142, and \agel0102, deflectors' observed stellar velocity dispersions are consistent with the model predicted velocity dispersions within their $1\sigma$ uncertainty bounds.
The $\sigma_{\rm obs}$ and $\sigma_{\rm pred}$ for AGEL0537 are consistent within the $2 \, \sigma$ uncertainty bound of $\sigma_{\rm obs}$. 
Thus, the lens model obtained for the above systems conforms with their true mass distribution.

For \agel2335, $\sigma_{\rm pred}=426\pm15 \, \rm km/s$ is much higher than the measured value $\sigma_{\rm obs}=321\pm27 \,\rm km/s$. 
In fact, $\sigma_{\rm pred}$ for \agel2335 is at the higher end of central velocity dispersions for ultra-massive quiescent galaxies \citep[e.g.,][]{Forrest:Wilson:2022}. 
The reason behind high $\sigma_{\rm pred}$ for \agel2335 may be linked with its large Einstein radius $R_{\rm Eins}=3\arcsec.59$, which is more typical of a galaxy group rather than an individual galaxy. 

In the DESI survey viewer, we find that the deflector galaxy in \agel2335 is the brightest among nearby galaxies at similar photometric redshift. Therefore, the deflector in \agel2335 may be a Brightest Cluster Galaxy (BCG), and the cluster halo may be contributing to its large Einstein radius.
Similarly, \agel0142, which also has a high $\sigma_{\rm pred}=385\pm29\,\rm km/s$, may be a BCG or Brightest Group Galaxy (BGG). 
Athough, for \agel0142 $\sigma_{\rm pred}$ is marginally consistent with $\sigma_{\rm obs}=316\pm40\,\rm km/s$ at $1\sigma$ uncertainty level and $R_{\rm Eins}=2\arcsec.33$ is typical of a galaxy-scale lens.
We have encircled \agel2335 and \agel0142 deflectors in our $\gamma$--$z$ diagrams discussed in the next section.

The cluster halo can affect the main deflector galaxy's density profile, including the slope parameter  \citep{Newman:Ellis:2015}.  Thus, when measuring the evolution of the density profile, the brightest cluster/group galaxy lenses should be distinguished from the isolated  galaxy lenses based on their host halo mass. 
The BCGs and BGGs may be outliers or have a different $\gamma$--$z$ trend compared to the isolated galaxies.
Because our current sample is limited, we do not exclude the above two lenses; however, future studies with combined past and current lens samples and known lens environments will enable this investigation. 

\subsection{$\gamma$--$z$ diagram}
\label{Result:gamma--z}
\begin{figure*}
\begin{center}
\includegraphics[clip=true,trim= 01mm 01mm 01mm 01mm,width=  1.0\textwidth]{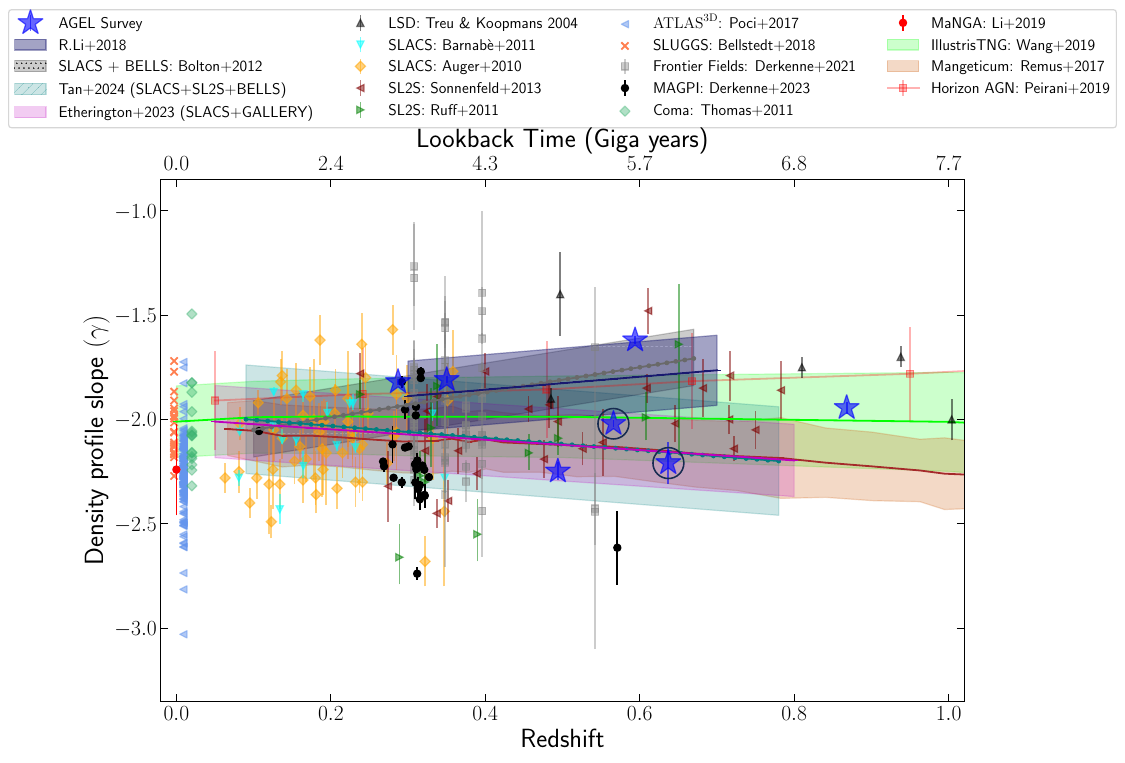}
\caption{Distribution of total mass density power-law profile slope with redshift as seen using lensing observations, dynamical observations, and simulations. 
\agel \, lenses modeled in this work are marked using blue stars. Encircled blue stars are the lenses \agel2335 and \agel1402 with high $\sigma_{\rm pred}$ (see Section \ref{Result: velocity dispersion}).
Markers for the data from various studies presented in this figure are explained in the legend and results from these studies are summarised in Table  \ref{tab:comparison}. 
All the shaded regions denote $68\%$ population scatter about the corresponding best-fit relation.
Overall, the $\gamma$--$z$ diagram has a non-uniform distribution with redshift, such that there are more observed data points at $z\lesssim 0.5$ than at higher redshift.
At $z\lesssim0.5$, the $\gamma$ versus $z$ distribution from lensing and dynamical observations overlaps well with that of simulations about the isothermal slope ($\gamma=-2$).
 However, at $z>0.5$, the $\gamma$--$z$ trends from 
 simulations (red squares, green, and brown shaded trends) and  some lensing observations (see gray, blue, magenta and teal shaded trends) seem to diverge.
\agel \, lenses with an average $\langle \gamma \rangle=-1.95\pm0.09$ and rms scatter of 0.21 are consistent with IllustrisTNG simulation; however, a larger lens sample is required to confirm the observed $\gamma$--$z$ relation at higher redshift. 
}
\label{fig:gam-z all}
\end{center}
\end{figure*}

Lens modeling provided us the slope parameter $\gamma= -\gamma^{\scaleto{\rm lens}{4pt}}$ for the total mass density profile of early type galaxy deflectors in our lens sample by fitting a power-law density profile upto the Einstein radius.
The best-fit $\gamma^{\scaleto{\rm lens}{4pt}}$ parameter along with $1\sigma$ uncertainty is provided in Table \ref{table:redshift}.
Our deflector galaxy sample of seven has an average (Lensing-Only) density profile slope of $\langle \gamma\rangle=-1.95 \pm 0.09$ with an rms scatter of $0.21$. 
Thus, the average slope of the density profile of deflector ETGs in our sample is consistent with an isothermal density profile ($\rho \propto r^{-2}$).

The distribution of density profile slope against redshift 
(and lookback time), the $\gamma$--$z$ diagram, is shown in Figure \ref{fig:gam-z all}.
The \agel \, deflectors are marked by blue stars. 
A linear Ordinary Least Squares, OLS($\gamma|z$), regression over our lens sample provided a slope of $d\langle \gamma \rangle/dz = -0.2\pm 0.2$ for the $\gamma$--$z$ relation, which is consistent with $\gamma$ not evolving with redshift. 
However, due to the small sample size, the $\gamma$--$z$ relation fit parameters are statistically insignificant with a low Pearson correlation coefficient r-value of $-0.2$.
Hence, we have not shown our line fit to the \agel \ lenses in the $\gamma$--$z$ diagram.

For comparison, in Figure \ref{fig:gam-z all} we also shows the $\gamma$--$z$ distribution/relation for ETGs obtained in the past lensing observations, dynamical observations, and hydro-dynamical simulations; namely, 
Magneticum, Horizon-AGN and IllustrisTNG \citep[from][respectively]{Remus:Dolag:Naab:2017, Peirani:Sonnenfeld:Gavazzi:2019, Wang:Vogelsberger:2019}. 
The lensing observations are from  \citet{Treu:Koopmans:2004}, \citet{Auger:Treu:Bolton:2010}, \citet{Barnabe:Czoske:2011}, \citet{Ruff:Gavazzi:2011}, \citet{Bolton:Brownstein:2012}, \citet{Sonnenfeld:Gavazzi:2013}, \citet{Li:Shu:Wang:2018}, \citet{Etherington:Nightingale:2023} and \citet{Tan:Shajib:2023}.
The dynamical observations are from \citet{Thomas:Saglia:2011}, \citet{Poci:Cappellari:2017}, \citet{Bellstedt:Forbes:2018}, \citet{Li:Li:Shao:2019},  \citet{Derkenne:McDermid:2021}, and \citet{Derkenne:McDermid:Poci:2023}.
Their markers are listed in the Figure \ref{fig:gam-z all} legend.
A summary of specifications such as the method used to obtain the density profile, sample size, redshift range, radial range,  change in $\gamma$ with redshift ($d\langle \gamma \rangle/dz$), and the average slope ($\langle \gamma \rangle$) of the sample obtained in all above studies are provided in Table \ref{tab:comparison}.

Upon analysing the comprehensive $\gamma$--$z$ diagram, shown in Figure \ref{fig:gam-z all}, we note the following.
Below $z<0.5$, the distribution of $\gamma$ for ETGs from individual observations \citep[e.g.,][]{Auger:Treu:Bolton:2010,Bolton:Brownstein:2012,  Derkenne:McDermid:2021,Tan:Shajib:2023} and simulations roughly overlap with each other about the isothermal slope ($\gamma=-2$); although, there is a significant scatter along the $\gamma$-axis.
In comparison, at $z \gtrsim 0.5$ there are diverging trends suggesting increasing ($d \langle \gamma \rangle/dz > 0$), decreasing ($d \langle \gamma \rangle/dz < 0$), and constant ($d \langle \gamma \rangle/dz \sim 0$) growth of density profile slope with redshift.  
Horizon-AGN simulation and lensing observations from \citet{Bolton:Brownstein:2012} and \citet{Li:Shu:Wang:2018}  suggests $d \langle \gamma \rangle/dz > 0$.
Magneticum simulation and lensing-only  analyses  from \citet{Etherington:Nightingale:2023} and \citet{Tan:Shajib:2023} suggest $d \langle \gamma \rangle/dz < 0$.  
IllustrisTNG simulation and lensing observations from \citet{Treu:Koopmans:2004}, \citet{Sonnenfeld:Treu:Gavazzi:2013}, and \agel \,  lenses suggest $d \langle \gamma \rangle/dz \sim 0$.
This indicates apparent conflicts within and between simulations and observations, especially at $z\gtrsim 0.5$.

Establishing the observed $\gamma$--$z$ trend, especially at $z\gtrsim 0.5$, is important to identify the physical processes responsible for galaxy growth by determining the simulations that best match the observation.
As discussed in the previous paragraph, the nature of the $\gamma$--$z$ relation at $z\gtrsim 0.5$ is unclear. 
Also, there are fewer observed data points with increasing redshift due to observational challenges at $z\gtrsim 0.5$. 
Building on this work, the addition of galaxy-scale lenses, especially at $z\gtrsim 0.5$, available in the \agel \, survey will help constrain the nature of the $\gamma$--$z$ relation. 
Quadrupling the current sample will reduce the uncertainty on overall average density profile slope by a factor of two.
Simultaneously, this will reduce the uncertainty in $\langle \gamma \rangle$ in each smaller redshift bin, thereby, strengthening the measured $\gamma$--$z$ relation.

Importantly, combining the quadrupled \agel \ lens sample with the past lensing observations across various redshift ranges (see Table \ref{tab:comparison}) can establish the $\gamma$--$z$ relation.
However, as Figure \ref{fig:gam-z all} shows and also discussed in the next section, there are discrepancies among observations as well; therefore, we need to understand and account for the differences in various studies before combining them.
An alternative approach is re-analyzing the past lensing sample using the same method. 
As manual lens modeling is time-consuming, this will require automated modeling, e.g.,  \citet{Ertl:Schuldt:2023} for quasar lenses, \citet{Etherington:Nightingale:2023} and \citet{Tan:Shajib:2023} for galaxy--galaxy lenses.
However, the precision of automated modeling may be lower than that of interactive modeling due to peculiarities in individual lenses.


\section{Discussion }
\label{sec: Discussion}

\begin{figure}
\includegraphics[clip=true,trim= 1.5mm 02mm 00mm 00mm,width=  0.5\textwidth]{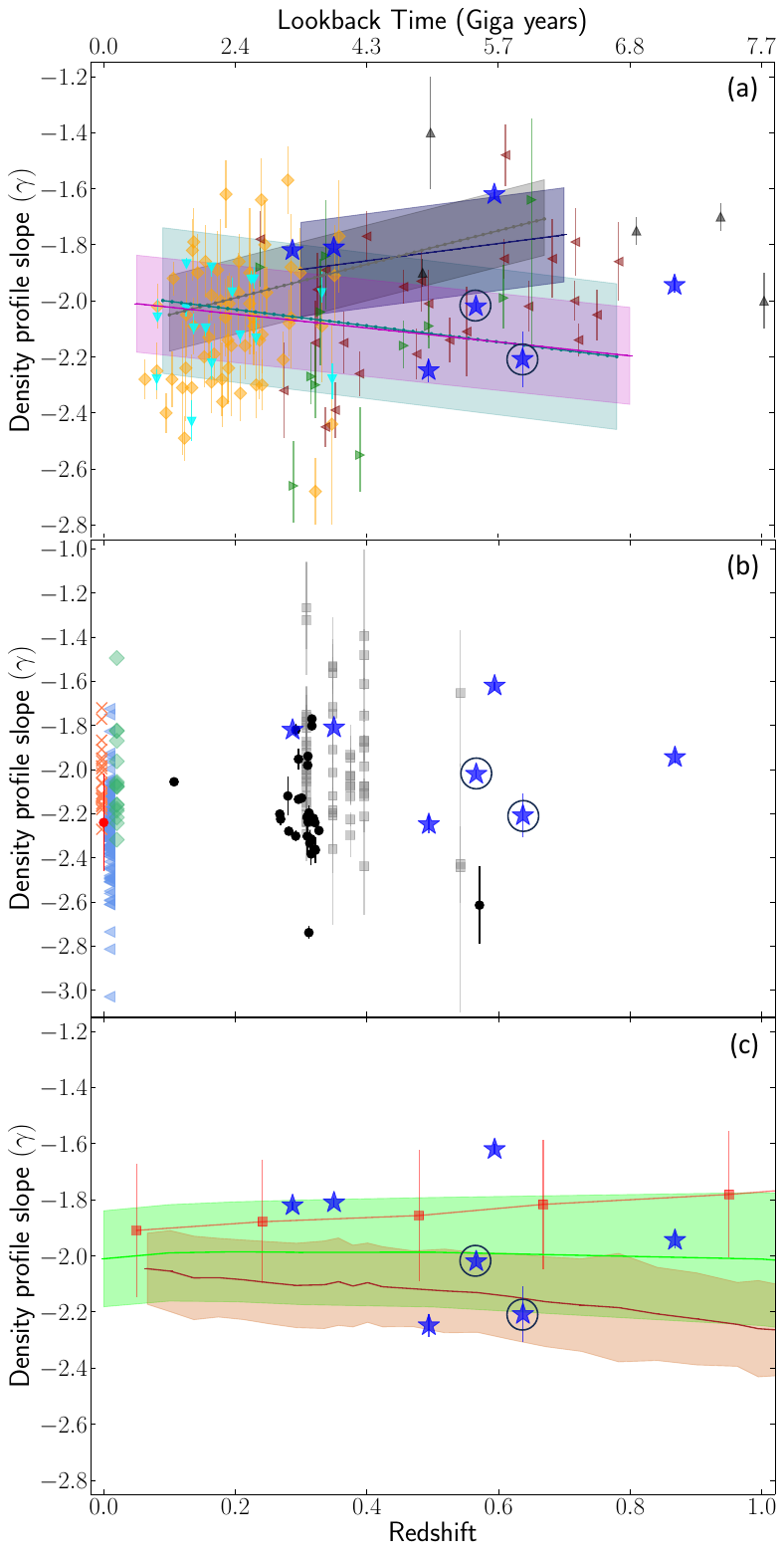}
\caption{A breakdown of $\gamma$-$z$ diagram into the results from lensing observations (top panel a), dynamical observations  (middle panel b), and simulations (bottom panel c). 
\agel \, lens sample modeled here is shown in all three panels for comparison with simulations and dynamical observations. Markers are the same as in Figure \ref{fig:gam-z all}.}
\label{fig:gam-z breakdown}
\end{figure}

\begin{table*}
    \centering
    \begin{tabular}{lllrrrl}
    \hline
         Study&  Method &  Sample size & redshift & $d\langle\gamma \rangle/dz$&  $\langle\gamma \rangle_{\rm sample}$  &  radial range\\
         (1) & (2) & (3) & (4) & (5) & (6) & (7) \\
         \hline
         This study  &  Lensing-only & 7 (\agel) & 0.3-0.9  & -0.2$\pm$ 0.2 & -1.95$\pm$0.09 & at $R_{\rm Eins}$ \\
         This study  &  L\&D\tablenotemark{a} & 7 (\agel) & 0.3-0.9  & $0.47\pm 0.17$ & -1.91$\pm$0.08 & $\sigma_{\rm obs}$ aperture size \\
         \hline
          \citet{Treu:Koopmans:2004} &  L\&D & 5 (LSD) & 0.5-0.94 & $\sim 0.0$ & -1.75$\pm$0.1 & $\leq R_{\rm eff}/8$ \\
          \citet{Auger:Treu:Bolton:2010}  & L\&D\tablenotemark{a} & 73 (SLACS) & 0.06-0.36 & $\sim 0.0$ & -2.08$\pm$0.04 & $\leq R_{\rm eff}/2$ \\
         \citet{Barnabe:Czoske:2011}  & L\&D & 16 (SLACS) & 0.08-0.33 & $\sim 0.0$ & -2.07$\pm$0.04 & $\leq R_{1/2}^{\rm 3D}$ \\
         \citet{Ruff:Gavazzi:2011}  &  L\&D\tablenotemark{a}  & 11 (SL2S) & 0.29-0.61 & $\sim 0.0$ & -2.22$\pm$0.19 & $\leq R_{\rm eff}/2$ \\
         \citet{Ruff:Gavazzi:2011}  &  L\&D  & 89 (SL2S,LSD, SLACS) & 0.06-0.94 & 0.25$\pm$0.11 & -2.12$\pm$0.04 & $\leq R_{\rm eff}/2$ \\
         \citet{Bolton:Brownstein:2012}  &  L\&D\tablenotemark{a}  & 79 (SLACS, BELLS) & 0.1-0.6  & 0.60$\pm$0.15 & - & $\leq R_{\rm eff}$ \\
         \citet{Sonnenfeld:Treu:Gavazzi:2013}  &  L\&D  & 25 (SL2S) & 0.2-0.8 & 0.10$\pm$0.12 & - & $\leq R_{\rm eff}/2$\\
         \citet{Li:Shu:Wang:2018} & L\&D  & 63 (BELLS, BELLS-  & 0.3-0.7 & 0.31$\pm$0.16 & -2.00$\pm$0.03 & $\leq 1 \arcsec$ \\
          &   &  GALLERY, SL2S)  &  &  &  & \\
         \citet{Etherington:Nightingale:2023}& Lensing-only\tablenotemark{b}  & 48 (SLACS, GALLERY) & 0.05-0.8 & -0.25$\pm$0.17 & -2.08$\pm$0.02 & at $R_{\rm Eins}$ \\
         \citet{Tan:Shajib:2023}& Lensing-only\tablenotemark{b}  & 77 (SLACS, SL2S, BELLS) & 0.09-0.78 & -0.29$\pm$0.18 & -2.05$\pm$0.04 & at $R_{\rm Eins}$ \\
         \hline
         \citet{Thomas:Saglia:2011} & Dynamical & 17 (Coma cluster) & $0.0231$ & - & -2.00$\pm$0.06 & $0.1 R_{\rm eff}$-$2 R_{\rm eff}$ \\
         \citet{Poci:Cappellari:2017} & Dynamical & 150 ($\rm ATLAS^{\rm 3D}$) & $\sim 0$ & - & -2.25$\pm$0.02 & $0.1 R_{\rm eff}$-$2 R_{\rm eff}$ \\
         \citet{Bellstedt:Forbes:2018} & Dynamical & 22 (SLUGGS) & $\sim 0$ & - & -2.06$\pm$0.04 & $0.1 R_{\rm eff}$-$2 R_{\rm eff}$\\
        \citet{Li:Li:Shao:2019} &  Dynamical & 2110 (MaNGA) & $\sim 0$ & - & -2.22$\pm$0.01 & $< R_{\rm eff}$\\
        \citet{Derkenne:McDermid:2021}  &  Dynamical &  64 (Frontier Fields cluster) & 0.29-0.55 & - & -2.01$\pm$0.04 & $0.1 R_{\rm eff}$-$2 R_{\rm eff}$ \\
        \citet{Derkenne:McDermid:Poci:2023}& Dynamical  & 28 (MAGPI)  & $\sim 0.31$ & - & -2.22$\pm$0.05 & $0.1 R_{\rm eff}$-$2 R_{\rm eff}$ \\
        \hline
        \citet{Remus:Dolag:Naab:2017} & Simulation  & Magneticum  & 0-1 & -0.21 & -2.13$\pm$0.15 & 0.4$R_{1/2}^{\rm 3D}$-4$R_{1/2}^{\rm 3D}$ \\
        \citet{Peirani:Sonnenfeld:Gavazzi:2019} & Simulation  & Horizon-AGN  & 0-1  & $>0$ & -1.85$\pm$0.23 & $0.5 R_{\rm eff}$-$R_{\rm eff}$ \\
        \citet{Wang:Vogelsberger:2019} & Simulation  & IllustrisTNG  & 0-1 & $\sim 0.0$ & -2.00$\pm$0.19  & 0.4$ R_{1/2}^{\rm 3D}$-4$R_{1/2}^{\rm 3D}$\\
        \hline
    \end{tabular}
    \caption{Details of studies presented in Figures \ref{fig:gam-z all} and \ref{fig:gam-z breakdown}. 
    Columns: (1) name of the study, (2) method used to constrain the galaxy total mass density profile, (3) sample size and the data survey, (4) redshift range of the deflectors/galaxies in the sample, (5) measured evolution of density profile slope with redshift (note that we represent the density profile as $\rho \propto r^{\gamma}$ with $\gamma <0$ in this paper), the trends from \citet{Ruff:Gavazzi:2011}, \citet{Li:Shu:Wang:2018}, and \citet{Tan:Shajib:2023} are $\partial \langle \gamma \rangle/\partial z $ (6)  average density profile slope of the sample, and (7) radial range over which slope ($\gamma$) was measured, for L\&D analysis this represents the aperture of $\sigma_{\rm obs}$ used.
    Abbreviation for various data surveys are: LSD- Lenses Structure and Dynamics \citep{Koopmans:Treu:2004}, SLACS- Sloan Lens ACS Survey \citep{Bolton:Burles:2006}, BELLS- BOSS Emission-Line Lens Survey \citep{Brownstein:Bolton:2012}, GALLERY- GALaxy-Ly$\alpha$ EmitteR sYstems Survey \citep{Shu:Bolton:2016}, SL2S- Strong Lensing Legacy Survey \citep{Cabanac:Alard:2007}, MAGPI- Middle Ages Galaxy Properties
with Integral field spectroscopy \citep{Foster:Mendel:2021}, MaNGA- Mapping Nearby Galaxies at Apache Point Observatory \citep{Bundy:Bershady:2015}, SLUGGS- SAGES Legacy
Unifying Globulars and GalaxieS Survey \citep{Brodie:Romanowsky:2014}. The radius $R^{3D}_{1/2}$ represents the (three-dimensional) half-mass radius. 
    }
    \tablenotetext{a}{Assumed isotropic orbits during spherical Jeans modeling to constrain $\gamma$ using $\sigma_{\rm obs}$.}
    \tablenotetext{b}{\citet{Etherington:Nightingale:2023} and \citet{Tan:Shajib:2023} used automated modeling.}
    \label{tab:comparison}
\end{table*}


The lensing observations, dynamical observations, and simulations use different types of data and methods to measure density profiles.
Here, we discuss the $\gamma$--$z$ relations individually from lensing observations, dynamical observations, and simulations and possible reasons behind discrepancies between and within these categories.
We have divided Figure \ref{fig:gam-z all} into the above three categories shown in panels (a), (b), and (c), respectively of Figure \ref{fig:gam-z breakdown}.
We have presented \agel \, lenses modeled here in all three panels for comparison.

\subsection{$\gamma$--$z$ relations from lensing observations}
\label{subsec: lensing}
 Even when comparing the same type of observations, it is essential to check  whether their $\gamma$ measurements represent the same property of the deflector mass density profile. 
 Table \ref{tab:comparison} indicates these differences for lensing studies shown in Panel (a) of Figure \ref{fig:gam-z breakdown}. 
 There are two types of lensing observations: ``Lensing-only" and joint ``Lensing and Dynamical" (L\&D) analysis.

The ``Lensing-only" method, described in Section \ref{Sec:modelling}, mainly requires a high-resolution lens image and directly fits a power-law mass model to the lens configuration up to $R_{\rm Eins}$ via the lens equation. 
This method provides us with complete lensing information,  including the constant power-law slope, $\gamma$, that is sensitive to the local slope at $R_{\rm Eins}$ \citep[see][]{Treu:2010}. 
The lens model posterior from this method can be affected by MSD (section \ref{subsec: predicting velcocity dispersion}), so it is essential to test the model's predicted velocity dispersion against the observed one.
In Panel (a) of Figure \ref{fig:gam-z breakdown}, this method is only used in \citet[][shown by magenta line]{Etherington:Nightingale:2023}, \citet[][shown by teal dotted line]{Tan:Shajib:2023}, and this work for the \agel \ lenses marked by blue stars. 


L\&D analysis uses the lens image only to obtain $R_{\rm Eins}$ and deflector light profile and requires source and deflector redshifts to measure projected mass enclosed within $R_{\rm Eins}$. It further constrains the spherical power-law density profile using independently measured stellar velocity dispersion and assumed stellar anisotropy through spherical Jeans modeling \citep[e.g.,][]{Koopmans:Treu:Bolton:2006}.
The choice of anisotropy affects the inferred density profile, and to remove associated systematics, a telescope-time-expensive resolved kinematics is required \citep{Cappellari:2016,  Shajib:Mozumdar:2023}.
The L\&D method essentially provides a global mass-weighted density profile slope within the radial range of $\sigma_{\rm obs}$ measurement \citep[see][]{Dutton:Treu:2014}, and it is a good proxy for average slope within $R_{\rm eff}$ \citep{Sonnenfeld:Treu:Gavazzi:2013}.


For a comparison, we also applied L\&D method assuming the same power-law model (Equation \ref{eqn: SPEMD convergence}) as used for Lensing-Only analysis for our sample. 
We used the Einstein radius and lens light profile already measured in our modeling process and used $\sigma_{\rm obs}$ to constrain the L\&D slopes ($\gamma^{\rm L\&D}$) by solving spherical Jeans equation using \textsc{Lenstronomy} \textsc{Galkin} routine (see Section \ref{subsec: predicting velcocity dispersion}).
We considered simple isotropic orbits for our ETG  lenses for this test. 

We find a sample average of $\langle \gamma^{\rm L\&D} \rangle = -1.91\pm0.08$ with an rms scatter of 0.19, which is  consistent with lensing-only measurements (cf. $\langle \gamma \rangle = -1.95\pm0.09$, scatter=0.21) within $1\sigma$ uncertainty level.
L\&D slopes for individual lenses are provided in Table \ref{table:redshift}. 
A one-to-one comparison between L\&D and lensing-only slopes is shown in Appendix Figure \ref{fig:gamma-gamma}.
We find that the two type of slopes relate as  $\gamma^{\rm L\&D} = (0.81\pm 0.12) \,(\gamma^{\rm \tiny{Lensing-only}} +2) - (1.94\pm 0.07)$ with a moderate Pearson coefficient (\textit{r}-value) of 0.55. Here, the intercept $- 1.94\pm 0.07 $ represents the mean value and scatter of $\gamma^{\rm L\&D}$ at the reference lensing-only  slope $\gamma^{\rm Lensing-only}=-2$.

A $\gamma$ versus $z$ plot with L\&D slopes for the \agel \ sample is shown in Appendix Figure \ref{fig:gamma-z_All_LnD}.
Interestingly, with redshift, L\&D slopes correlate as  $\gamma^{\rm L\&D}=(0.47\pm 0.17) \, (z-0.57) - (1.89\pm0.06)$ with Pearson r-value of 0.53. 
This is consistent with past L\&D observations but contrary to recent lensing-only observations \citep{Etherington:Nightingale:2023, Tan:Shajib:2023}, whereas our lensing-only measurements suggest no slope evolution with $d\gamma/dz$ consistent with zero (see Table \ref{tab:comparison}). 
Nonetheless, we emphasize the need to use a larger sample in the future to test the above correlations.

If the assumption of the power-law mass model is correct, there should be no discrepancies between Lensing-only and L\&D methods except if there are systematics in $\sigma_{\rm obs}$ measurement or the  anisotropy assumption used in the L\&D analysis. 
These two factors can bias the $\gamma$ values and add scatter in the $\gamma$--$z$ distribution \citep{Xu:Springel:Sluse:2017, Birrer:Shajib:Galan:2020}.
On the other hand, the lensing-only and L\&D methods will provide different slope values if the actual density profile is more complex than a simple power-law. 
Either or both of the above could be a reason why the lensing-only $\gamma$--$z$ trends from \citet{Etherington:Nightingale:2023}, \citet{Tan:Shajib:2023}, and this work are different from L\&D studies, e.g., \citet[][gray line]{Bolton:Brownstein:2012} and \citet[][navy-blue line]{Li:Shu:Wang:2018}.

\citet{Etherington:Nightingale:2023} also find a higher scatter in $\gamma$ measured via the lensing-only method than the L\&D method for the same lens sample.
They argue that  $R_{\rm Eins}$ is close to an inflection point beyond which dark matter starts to dominate the baryonic matter, and the total density profile deviates from the power-law. 
They suggest that the lensing-only measurements are, therefore, sensitive to the location of the inflection point, enhancing rms scatter in the measured slopes. L\&D slopes, on the other hand, are averaged over an extended radial range, resulting in a smaller population scatter.
If true, this could be behind the higher population scatter about the $\gamma$--$z$ trend from \citet[][scatter=0.26]{Tan:Shajib:2023} compared to past L\&D studies \citep[][scatter=0.14]{Bolton:Brownstein:2012}.

In case the true mass model is not a power-law, the difference in the radial range over which slope is weighted in the L\&D observations or the difference in $\sigma_{\rm obs}$ aperture  can produce different $\gamma$ values even for the same lens.  
Thus, the use of different radial ranges in the past L\&D observations \citep[e.g., see][]{Auger:Treu:Bolton:2010,  Bolton:Brownstein:2012, Li:Shu:Wang:2018} may be why 
their distributions do not align well and add a significant scatter in the $\gamma$--$z$ diagram.
Some studies are testing the possibility of a flexible non-power-law lens mass model \citep[e.g.,][]{Birrer:Shajib:Galan:2020, Kochanek:2020, Shajib:Treu:Birrer:2021}.
For example, for future studies with a complete sample,  \citet{Tan:Shajib:2023} suggest using a composite mass model with dark and baryonic matter modeled separately in the L\&D analysis to obtain the total density profile for a better comparison with simulations.



\subsection{$\gamma$--$z$ relations from 2D dynamical observations}
\label{subsec: dynamical}
Many 2D dynamical studies were restricted to local ETGs at $z\sim 0$ some years ago  \citep[e.g.,][as shown in  panel b of Figure \ref{fig:gam-z breakdown}]{Poci:Cappellari:2017, Bellstedt:Forbes:2018, Li:Li:Shao:2019, Thomas:Saglia:2011}.
Improved instrumentation is now pushing the resolved kinematic observation limit to intermediate redshifts $\sim 0.3<z<0.55$, such as \citet[][black circles]{Derkenne:McDermid:2021} and  \citet[][gray squares]{Derkenne:McDermid:Poci:2023}, shown in Panel (b) of Figure \ref{fig:gam-z breakdown}.

Using 2D stellar kinematics and high-resolution HST images, dynamical observations measure the spatial total density profile via stellar dynamical modeling (Jeans Anisotropic Multi-Gaussian Expansion or Schwarzschild modeling) with unconstrained anisotropy and inclination angle.
These studies calculate $\gamma$ as the mean logarithmic slope ($\Delta \log \rho/\Delta \log r$) over a certain radial range.
Thus, using different radial ranges will produce different  mean logarithmic slope values if the galaxy's total mass profile is not a power-law with a constant slope. 

\citet{Derkenne:McDermid:Poci:2023} dynamically 
measured the mean logarithmic slope of density profile over the same radial range of $\rm 0.1 \, R_{\rm eff}$ to $\rm 2 \, R_{\rm eff}$ for 28 MAGPI survey galaxies.
Past dynamical studies have used variable radial ranges to measure $\gamma$. 
Therefore, for better comparison,  \citet{Derkenne:McDermid:Poci:2023} re-calculated slopes for the past ETG sample, except for \citet{Li:Li:Shao:2019}, using the same radial range $\rm 0.1 - 2\, R_{\rm eff}$. 
Figures \ref{fig:gam-z all} and \ref{fig:gam-z breakdown} show the updated slopes for the past dynamical sample from \citet{Derkenne:McDermid:Poci:2023} .

As summarised in Table \ref{tab:comparison}, the sample average density profile slopes obtained in dynamical studies at $z\sim0$ and $0.3<z<0.55$ are close to the isothermal slope; therefore, \citet{Derkenne:McDermid:2021} and \citet{Derkenne:McDermid:Poci:2023} suggest an overall lack of evolution (i.e., $d\langle\gamma \rangle/dz \sim 0$) of the mass density profile slope with redshift for ETGs.
However, this conclusion requires further investigation because there is still a lack of resolved kinematic data between $0<z<0.3$ (see Panel b), and slope measurements for some dynamical samples have very high uncertainty \citep[e.g., 0.2 to 0.6 in][]{Derkenne:McDermid:2021}.

The lack of dynamical observations between $0<z<0.3$ is supplemented by the SLACS lens sample for which \citet{Auger:Treu:Bolton:2010} and \citet{Bolton:Brownstein:2012} find a shallower sample average slope ($2.08\pm0.04$) using L\&D method than the $\rm ATLAS^{\rm 3D}$, MaNGA and MAGPI dynamical observations ($\langle\gamma \rangle \sim -2.23\pm 0.02$). 
This could be due to slopes weighted/averaged over different radial ranges (see Table \ref{tab:comparison}) for possible   density profiles with varying slopes.
 
The difficulty in acquiring spatially resolved kinematics at high redshifts significantly contributes to the large uncertainties in the dynamical slope measurements. 
This is where \agel \, lenses are highly complementary and can provide total density profile slope measurement for galaxy lenses at high-$z$ only using high-quality photometric lens images.
In the future, a side-by-side comparison by applying lensing and 2D dynamical analysis on the same dataset will be insightful for combining results from two types of observations to refine the $\gamma$--$z$ relation.

\subsection{ $\gamma$--$z$ relations from simulations}
\label{subsec: simulation} 

Comparison with observations confirms the simulation results and lets us understand galaxy evolution processes in more detail than observationally possible.
Panel (c) in Figure \ref{fig:gam-z breakdown} shows that at $z\lesssim 1$,  Magneticum, IllustrisTNG, and Horizon-AGN simulations suggest different evolution of ETG density profile slope, $\gamma$, with redshift.
Though at $z\lesssim 0.5$, the $\gamma$--$z$ relations from Magneticum and Horizon-AGN simulation seem to overlap well with the IllustrisTNG simulation ($\gamma\sim -2$) within their $1\sigma$ scatter bounds. 
All three simulated trends diverge as we move towards higher redshift. 

Figures \ref{fig:gam-z all} and \ref{fig:gam-z breakdown}  represent the simulated $\gamma$--$z$ relation only until $z=1$ because the observational data is limited to $z\sim 1$.  
Magneticum simulation found a monotonous shallowing of $\gamma$ for ETGs with decreasing redshift, starting from $\gamma \sim -3$ at $z\sim 2$ to  $\gamma \sim -2$ at $z\sim 0$ \citep{Remus:Dolag:Naab:2017}.
In contrast, Horizon-AGN simulation found a continuous  steepening of $\gamma$ with decreasing redshift, starting from $\gamma \sim -1.6$ at $z=2$ to $\gamma \sim -2$ at $z\sim 0$ \citep{Peirani:Sonnenfeld:Gavazzi:2019}.
IllustrisTNG simulation, on the other hand, found a multi-phase $\gamma$--$z$ trend for ETGs \citep{Wang:Vogelsberger:2019}. 
IllustrisTNG simulation found that, for the main progenitor branch galaxies, $\gamma$ steepens up to $-2.2$ during $2<z<4$ due to gas-rich evolution, then shallows down to $\gamma \sim -2$ between $1<z<2$ mainly due to AGN feedback and further the isothermal slope ($\gamma \sim -2$) is maintained with a passive gas-poor evolution at $z \lesssim 1$.

The reason behind different $\gamma$--$z$ simulation trends could be different models for galaxy evolution processes, e.g., baryonic versus dark matter interactions, gas cooling, subgrid physics, stellar feedback, or AGN feedback,    adopted in simulations \citep{Peirani:Sonnenfeld:Gavazzi:2019, Wang:Vogelsberger:2020, Mukherjee:Koopmans:2021}. 
Additionally, the discrepancy between simulations and lensing observations may be due to challenges in comparing the two.
One challenge is due to methodological differences \citep{Remus:Dolag:Naab:2017}, e.g., the use of spatial (3D) radius in simulations rather than projected radius as in lensing observations when calculating $\gamma$ (see Table \ref{tab:comparison}).
Secondly, simulations can track the evolution of the same galaxy sample over cosmic time, which is impractical for observations. 
\citet{Filipp:Shu:2023} present an approach for direct comparison between observations and simulations by identifying direct counterparts of galaxies in observations and simulations. However, the current sample of lenses needs to be expanded to apply their strategy.

The lensing-only average $\langle \gamma \rangle=-1.95\pm0.09 $ and $d\langle \gamma \rangle/dz = -0.2\pm 0.2$  for \agel \, sample modeled in this work seems consistent with the IllustrisTNG simulation with $d\langle \gamma \rangle/dz \sim 0$ at $z \lesssim 1$. 
However, it is not consistent with other simulations and some observations that suggest $d\langle \gamma \rangle/dz >0$ (e.g., \citet{Li:Shu:Wang:2018}, \citet{Bolton:Brownstein:2012}, Horizon-AGN) or $d\langle \gamma \rangle/dz <0$ (e.g., \citet{Etherington:Nightingale:2023}, Magneticum). 
On the other hand, the L\&D analysis results for \agel \, sample ($\langle \gamma \rangle=-1.91\pm0.08 $ and $d\langle \gamma \rangle/dz = 0.47\pm 0.17$)  
is consistent with the past L\&D observations and Horizon-AGN simulation with $d\langle \gamma \rangle/dz >0$.
Distinguishing which simulation is favored by observations is possible only after establishing the observed  $\gamma$--$z$ trend at $z \gtrsim 0.5$ and accounting for possible methodological differences among observations and simulations.
In future studies, the addition of high-$z$ \agel \, lenses and, further, careful combination with past lensing observations e.g., SL2S \citep[][$0.2 \lesssim z \lesssim 0.8$]{Sonnenfeld:Treu:Gavazzi:2013}, BELLS \citep[][$0.4\lesssim z \lesssim0.7$]{Brownstein:Bolton:2012}, and BELLS-GALLERY \citep[][$0.4\lesssim z \lesssim0.7$]{Shu:Bolton:2016} samples can help constrain the nature of $\gamma$--$z$ relation at $z \gtrsim 0.5$.


\section{Conclusion}
\label{Sec:conclusions}
The observed evolution of galaxy density profiles with cosmic time is a critical test of theoretical models for galaxy evolution.  However, density profiles are increasingly challenging to measure at higher redshifts.  Here, we measure the total matter density profile slope ($\gamma$) versus redshift ($z$) relation using strong gravitational lenses at \zdefl$>0.3$, which complement the dynamical observations at lower redshifts.
In the $\gamma$--$z$ diagram, we add seven new galaxy-scale lenses with $0.3<$ \zdefl $<0.9$ from the \astrod\ Galaxy Evolution with Lenses (\agel) survey  (Section \ref{Sec:Data}).
We also investigate the apparent conflict between lensing observations, purely dynamical observations, and simulations
in the $\gamma$--$z$ diagram.

To obtain the total mass density profile of deflector galaxies, we perform lensing-only modeling using high-resolution HST images.  We use state-of-the-art lens modeling software GLEE and assume a power-law mass profile ($\rho \propto r^{\gamma}$) with a constant slope ($\gamma <0$) for deflector galaxies (Section \ref{Sec:modelling}).
To provide an independent test of our lens mass models, we also measure the stellar velocity dispersion, $\sigma_{\rm obs}$, for deflectors in four lenses and obtained $\sigma_{\rm obs}$ for the remaining three lenses from SDSS-BOSS survey \citep{Thomas:Steele:2013}. 

We successfully reconstruct the lensing configuration and the background source surface brightness for all seven lenses (see Figures \ref{fig:models} and \ref{fig:models appendix}).  Similar to existing studies, all of the seven deflectors in our pilot sample are Early Type Galaxies (ETGs).
We find that the velocity dispersions , $\sigma_{\rm pred}$, predicted by our lens model for the deflector ETGs are consistent with the observed velocity dispersions, suggesting that our lens models are accurate, for all lenses except for \agel2335 (Section \ref{Sec:Results}).
The deflectors in \agel2335 and \agel0142 are likely to be the brightest cluster/group galaxies as both have high $\sigma_{\rm pred}$ potentially due to the cluster/group  halo contributing to the lensing potential. 
We suggest future studies with larger sample to separate the brightest cluster/group galaxies from isolated galaxies when studying the evolution of total density profile.


The deflector galaxies in our pilot sample have an average lensing-only density profile slope of $-1.95\pm0.09$ with an rms scatter of $0.21$, consistent with an isothermal density profile.
The $\gamma$--$z$ distribution of our sample with $d\langle \gamma \rangle/dz = -0.2\pm 0.2$ is broadly consistent with the IllustrisTNG simulation \citep{Wang:Vogelsberger:2019}, dynamical observations \citep{Derkenne:McDermid:2021,  Derkenne:McDermid:Poci:2023} and some lensing observations \citep{Treu:Koopmans:2004,  Sonnenfeld:Treu:Gavazzi:2013}, that suggest no slope evolution with redshift (Section \ref{Sec:Results}). 
The lack of evolution in $\gamma$ at $z\lesssim 1$ is consistent with the predictions of ETGs passively evolving through gas-poor mergers and maintaining a shallow isothermal ($\gamma=-2$) density profile \citep{Wang:Vogelsberger:2019}.
However, we need a larger statistical sample to confirm our findings.

We also present density profile slopes measured using the 
 joint lensing and dynamical (L\&D) analysis. 
 The L\&D method, assuming isotropic stellar orbits, provides a sample average slope of $-1.91\pm0.08$ with an rms scatter of 0.19, consistent with our lensing-only analysis.
Regarding evolution with redshift, the L\&D slopes tend to steepen with decreasing redshift, consistent with some past L\&D observations \citep{Bolton:Brownstein:2012, Li:Shu:Wang:2018} but contrary to recent lensing-only observations \citep{Etherington:Nightingale:2023, Tan:Shajib:2023}. 
This finding requires further detailed testing with varying anisotropy and a larger lens sample.

In the updated $\gamma$--$z$ diagram, apart from differences between observations and simulations, we notice differences within observations \citep[e.g.,][]{Bolton:Brownstein:2012, Etherington:Nightingale:2023} and  within simulations (Section \ref{sec: Discussion}).
The differences between simulations are mainly due to the different physical processes adopted, such as the strength of AGN feedback \citep{Peirani:Sonnenfeld:Gavazzi:2019}.
Based on our comparisons, the possible reasons for the apparent discrepancies  with and within observations  
are a combination of the following:
1) Systematics in the assumed anisotropy profile or  $\sigma_{\rm obs}$ used in the joint lensing and dynamical (L\&D) and purely dynamical modeling; 
2) Simplistic assumption of a power-law mass profile; and 
3) Challenges in comparing observations with the simulations.
If the assumed power-law mass profile is inaccurate, using different methodologies such as lensing-only and L\&D to obtain the density profile or different radial ranges over which  $\gamma$ is averaged will provide different values.


Building on the analysis in this paper, we plan to quadruple the modeled \agel\ lens sample  available from the \agel \, survey, especially at $z>0.5$, in a follow-up paper.
This will help constrain the nature of the $\gamma$--$z$ relation at the high-redshift end by reducing uncertainty in $\langle\gamma \rangle$ by a factor of two.
Moreover, by combining our sample with past lensing observations after accounting for differences in the methodology we can refine the $\gamma$--$z$ relation even further.
We will use the expanded sample to simultaneously investigate the evolution in associated galaxy properties (e.g., size, mass, kinematics, and stellar mass density) with cosmic time, which will provide improved tests for galaxy evolution models.

    
\begin{acknowledgements}
This research was supported by the Australian
Research Council Centre of Excellence for All Sky Astrophysics in 3 Dimensions (ASTRO 3D), through project
number CE170100013.
Authors thank Yiping Shu for their valuable feedback on this manuscript. Authors thank Alessandro Sonnenfeld for their  suggestions regarding joint lensing and dynamical analysis.
Authors also thank the anonymous referee for their constructive comments that improved the clarity of this paper.
SHS and SE thank the Max Planck Society for support through the Max Planck Research Group and Max Planck Fellowship for SHS. 
AJS was supported by NASA through the NASA Hubble Fellowship grant HST-HF2-51492 from the Space Telescope Science Institute, which is operated by the Association of Universities for Research in Astronomy, Inc., for NASA, under contract NAS5-26555.
TJ and KVGC gratefully acknowledge financial support from NASA through grant HST-GO-16773, the Gordon and Betty Moore Foundation through Grant GBMF8549, the National Science Foundation through grant AST-2108515, and from a Dean’s Faculty Fellowship.
SMS acknowledges funding from the Australian Research Council (DE220100003).
S.L. acknowledges support by FONDECYT grant 1231187. 
\\
Other software used: \textsc{Numpy} \citep{NumPy:2020}, \textsc{Scipy} \citep{SciPy:2020}, \textsc{Astropy} \citep{Astropy:2013,  Astropy:2018, Astropy:2022}, \textsc{Matplotlib} \citep{matplotlib:2007}, \textsc{Multiprocess} \citep{Multiprocess:McKerns:Strand:2012}
\end{acknowledgements}

\bibliographystyle{aasjournal}
\bibliography{Lens_bibliography}

\begin{thebibliography}{}
\expandafter\ifx\csname natexlab\endcsname\relax\def\natexlab#1{#1}\fi
\providecommand{\url}[1]{\href{#1}{#1}}
\providecommand{\dodoi}[1]{doi:~\href{http://doi.org/#1}{\nolinkurl{#1}}}
\providecommand{\doeprint}[1]{\href{http://ascl.net/#1}{\nolinkurl{http://ascl.net/#1}}}
\providecommand{\doarXiv}[1]{\href{https://arxiv.org/abs/#1}{\nolinkurl{https://arxiv.org/abs/#1}}}

\bibitem[{{Akritas} \& {Bershady}(1996)}]{Akritas:Bershady:1996}
{Akritas}, M.~G., \& {Bershady}, M.~A. 1996, \apj, 470, 706, \dodoi{10.1086/177901}

\bibitem[{{Astropy Collaboration} {et~al.}(2013){Astropy Collaboration}, {Robitaille}, {Tollerud}, {Greenfield}, {Droettboom}, {Bray}, {Aldcroft}, {Davis}, {Ginsburg}, {Price-Whelan}, {Kerzendorf}, {Conley}, {Crighton}, {Barbary}, {Muna}, {Ferguson}, {Grollier}, {Parikh}, {Nair}, {Unther}, {Deil}, {Woillez}, {Conseil}, {Kramer}, {Turner}, {Singer}, {Fox}, {Weaver}, {Zabalza}, {Edwards}, {Azalee Bostroem}, {Burke}, {Casey}, {Crawford}, {Dencheva}, {Ely}, {Jenness}, {Labrie}, {Lim}, {Pierfederici}, {Pontzen}, {Ptak}, {Refsdal}, {Servillat}, \& {Streicher}}]{Astropy:2013}
{Astropy Collaboration}, {Robitaille}, T.~P., {Tollerud}, E.~J., {et~al.} 2013, \aap, 558, A33, \dodoi{10.1051/0004-6361/201322068}

\bibitem[{{Astropy Collaboration} {et~al.}(2018){Astropy Collaboration}, {Price-Whelan}, {Sip{\H{o}}cz}, {G{\"u}nther}, {Lim}, {Crawford}, {Conseil}, {Shupe}, {Craig}, {Dencheva}, {Ginsburg}, {VanderPlas}, {Bradley}, {P{\'e}rez-Su{\'a}rez}, {de Val-Borro}, {Aldcroft}, {Cruz}, {Robitaille}, {Tollerud}, {Ardelean}, {Babej}, {Bach}, {Bachetti}, {Bakanov}, {Bamford}, {Barentsen}, {Barmby}, {Baumbach}, {Berry}, {Biscani}, {Boquien}, {Bostroem}, {Bouma}, {Brammer}, {Bray}, {Breytenbach}, {Buddelmeijer}, {Burke}, {Calderone}, {Cano Rodr{\'\i}guez}, {Cara}, {Cardoso}, {Cheedella}, {Copin}, {Corrales}, {Crichton}, {D'Avella}, {Deil}, {Depagne}, {Dietrich}, {Donath}, {Droettboom}, {Earl}, {Erben}, {Fabbro}, {Ferreira}, {Finethy}, {Fox}, {Garrison}, {Gibbons}, {Goldstein}, {Gommers}, {Greco}, {Greenfield}, {Groener}, {Grollier}, {Hagen}, {Hirst}, {Homeier}, {Horton}, {Hosseinzadeh}, {Hu}, {Hunkeler}, {Ivezi{\'c}}, {Jain}, {Jenness}, {Kanarek}, {Kendrew}, {Kern}, {Kerzendorf}, {Khvalko}, {King}, {Kirkby}, {Kulkarni},
  {Kumar}, {Lee}, {Lenz}, {Littlefair}, {Ma}, {Macleod}, {Mastropietro}, {McCully}, {Montagnac}, {Morris}, {Mueller}, {Mumford}, {Muna}, {Murphy}, {Nelson}, {Nguyen}, {Ninan}, {N{\"o}the}, {Ogaz}, {Oh}, {Parejko}, {Parley}, {Pascual}, {Patil}, {Patil}, {Plunkett}, {Prochaska}, {Rastogi}, {Reddy Janga}, {Sabater}, {Sakurikar}, {Seifert}, {Sherbert}, {Sherwood-Taylor}, {Shih}, {Sick}, {Silbiger}, {Singanamalla}, {Singer}, {Sladen}, {Sooley}, {Sornarajah}, {Streicher}, {Teuben}, {Thomas}, {Tremblay}, {Turner}, {Terr{\'o}n}, {van Kerkwijk}, {de la Vega}, {Watkins}, {Weaver}, {Whitmore}, {Woillez}, {Zabalza}, \& {Astropy Contributors}}]{Astropy:2018}
{Astropy Collaboration}, {Price-Whelan}, A.~M., {Sip{\H{o}}cz}, B.~M., {et~al.} 2018, \aj, 156, 123, \dodoi{10.3847/1538-3881/aabc4f}

\bibitem[{{Astropy Collaboration} {et~al.}(2022){Astropy Collaboration}, {Price-Whelan}, {Lim}, {Earl}, {Starkman}, {Bradley}, {Shupe}, {Patil}, {Corrales}, {Brasseur}, {N{\"o}the}, {Donath}, {Tollerud}, {Morris}, {Ginsburg}, {Vaher}, {Weaver}, {Tocknell}, {Jamieson}, {van Kerkwijk}, {Robitaille}, {Merry}, {Bachetti}, {G{\"u}nther}, {Aldcroft}, {Alvarado-Montes}, {Archibald}, {B{\'o}di}, {Bapat}, {Barentsen}, {Baz{\'a}n}, {Biswas}, {Boquien}, {Burke}, {Cara}, {Cara}, {Conroy}, {Conseil}, {Craig}, {Cross}, {Cruz}, {D'Eugenio}, {Dencheva}, {Devillepoix}, {Dietrich}, {Eigenbrot}, {Erben}, {Ferreira}, {Foreman-Mackey}, {Fox}, {Freij}, {Garg}, {Geda}, {Glattly}, {Gondhalekar}, {Gordon}, {Grant}, {Greenfield}, {Groener}, {Guest}, {Gurovich}, {Handberg}, {Hart}, {Hatfield-Dodds}, {Homeier}, {Hosseinzadeh}, {Jenness}, {Jones}, {Joseph}, {Kalmbach}, {Karamehmetoglu}, {Ka{\l}uszy{\'n}ski}, {Kelley}, {Kern}, {Kerzendorf}, {Koch}, {Kulumani}, {Lee}, {Ly}, {Ma}, {MacBride}, {Maljaars}, {Muna}, {Murphy}, {Norman},
  {O'Steen}, {Oman}, {Pacifici}, {Pascual}, {Pascual-Granado}, {Patil}, {Perren}, {Pickering}, {Rastogi}, {Roulston}, {Ryan}, {Rykoff}, {Sabater}, {Sakurikar}, {Salgado}, {Sanghi}, {Saunders}, {Savchenko}, {Schwardt}, {Seifert-Eckert}, {Shih}, {Jain}, {Shukla}, {Sick}, {Simpson}, {Singanamalla}, {Singer}, {Singhal}, {Sinha}, {Sip{\H{o}}cz}, {Spitler}, {Stansby}, {Streicher}, {{\v{S}}umak}, {Swinbank}, {Taranu}, {Tewary}, {Tremblay}, {de Val-Borro}, {Van Kooten}, {Vasovi{\'c}}, {Verma}, {de Miranda Cardoso}, {Williams}, {Wilson}, {Winkel}, {Wood-Vasey}, {Xue}, {Yoachim}, {Zhang}, {Zonca}, \& {Astropy Project Contributors}}]{Astropy:2022}
{Astropy Collaboration}, {Price-Whelan}, A.~M., {Lim}, P.~L., {et~al.} 2022, \apj, 935, 167, \dodoi{10.3847/1538-4357/ac7c74}

\bibitem[{{Auger} {et~al.}(2010){Auger}, {Treu}, {Bolton}, {Gavazzi}, {Koopmans}, {Marshall}, {Moustakas}, \& {Burles}}]{Auger:Treu:Bolton:2010}
{Auger}, M.~W., {Treu}, T., {Bolton}, A.~S., {et~al.} 2010, \apj, 724, 511, \dodoi{10.1088/0004-637X/724/1/511}

\bibitem[{{Avila} {et~al.}(2015){Avila}, {Hack}, {Cara}, {Borncamp}, {Mack}, {Smith}, \& {Ubeda}}]{Avila:Hack:2015}
{Avila}, R.~J., {Hack}, W., {Cara}, M., {et~al.} 2015, in Astronomical Society of the Pacific Conference Series, Vol. 495, Astronomical Data Analysis Software an Systems XXIV (ADASS XXIV), ed. A.~R. {Taylor} \& E.~{Rosolowsky}, 281, \dodoi{10.48550/arXiv.1411.5605}

\bibitem[{{Barkana}(1998)}]{Barkana:1998}
{Barkana}, R. 1998, \apj, 502, 531, \dodoi{10.1086/305950}

\bibitem[{{Barnab{\`e}} {et~al.}(2011){Barnab{\`e}}, {Czoske}, {Koopmans}, {Treu}, \& {Bolton}}]{Barnabe:Czoske:2011}
{Barnab{\`e}}, M., {Czoske}, O., {Koopmans}, L. V.~E., {Treu}, T., \& {Bolton}, A.~S. 2011, \mnras, 415, 2215, \dodoi{10.1111/j.1365-2966.2011.18842.x}

\bibitem[{{Barnes} \& {Hernquist}(1991)}]{Barnes:Hernquist:1991}
{Barnes}, J.~E., \& {Hernquist}, L.~E. 1991, \apjl, 370, L65, \dodoi{10.1086/185978}

\bibitem[{{Bellstedt} {et~al.}(2018){Bellstedt}, {Forbes}, {Romanowsky}, {Remus}, {Stevens}, {Brodie}, {Poci}, {McDermid}, {Alabi}, {Chevalier}, {Adams}, {Ferr{\'e}-Mateu}, {Wasserman}, \& {Pandya}}]{Bellstedt:Forbes:2018}
{Bellstedt}, S., {Forbes}, D.~A., {Romanowsky}, A.~J., {et~al.} 2018, \mnras, 476, 4543, \dodoi{10.1093/mnras/sty456}

\bibitem[{{Binney}(1978)}]{Binney:1978}
{Binney}, J. 1978, \mnras, 183, 501, \dodoi{10.1093/mnras/183.3.501}

\bibitem[{{Binney} \& {Mamon}(1982)}]{Binney:Mamon:1982}
{Binney}, J., \& {Mamon}, G.~A. 1982, \mnras, 200, 361, \dodoi{10.1093/mnras/200.2.361}

\bibitem[{{Birrer}(2021)}]{Birrer:2021}
{Birrer}, S. 2021, \apj, 919, 38, \dodoi{10.3847/1538-4357/ac1108}

\bibitem[{{Birrer} \& {Amara}(2018)}]{Birrer:Amara:2018}
{Birrer}, S., \& {Amara}, A. 2018, Physics of the Dark Universe, 22, 189, \dodoi{10.1016/j.dark.2018.11.002}

\bibitem[{{Birrer} {et~al.}(2020){Birrer}, {Shajib}, {Galan}, {Millon}, {Treu}, {Agnello}, {Auger}, {Chen}, {Christensen}, {Collett}, {Courbin}, {Fassnacht}, {Koopmans}, {Marshall}, {Park}, {Rusu}, {Sluse}, {Spiniello}, {Suyu}, {Wagner-Carena}, {Wong}, {Barnab{\`e}}, {Bolton}, {Czoske}, {Ding}, {Frieman}, \& {Van de Vyvere}}]{Birrer:Shajib:Galan:2020}
{Birrer}, S., {Shajib}, A.~J., {Galan}, A., {et~al.} 2020, \aap, 643, A165, \dodoi{10.1051/0004-6361/202038861}

\bibitem[{Birrer {et~al.}(2021)Birrer, Shajib, Gilman, Galan, Aalbers, Millon, Morgan, Pagano, Park, Teodori, Tessore, Ueland, de~Vyvere, Wagner-Carena, Wempe, Yang, Ding, Schmidt, Sluse, Zhang, \& Amara}]{Birrer:Shajib:Gilman:2021}
Birrer, S., Shajib, A.~J., Gilman, D., {et~al.} 2021, Journal of Open Source Software, 6, 3283, \dodoi{10.21105/joss.03283}

\bibitem[{{Blandford} \& {Kochanek}(1987)}]{Blandford:Kochanek:1987}
{Blandford}, R.~D., \& {Kochanek}, C.~S. 1987, \apj, 321, 658, \dodoi{10.1086/165660}

\bibitem[{{Blandford} \& {Narayan}(1992)}]{Blandford:Narayan:1992}
{Blandford}, R.~D., \& {Narayan}, R. 1992, \araa, 30, 311, \dodoi{10.1146/annurev.astro.30.1.311}

\bibitem[{{Blumenthal} {et~al.}(1984){Blumenthal}, {Faber}, {Primack}, \& {Rees}}]{Blumenthal:Faber:1984}
{Blumenthal}, G.~R., {Faber}, S.~M., {Primack}, J.~R., \& {Rees}, M.~J. 1984, \nat, 311, 517, \dodoi{10.1038/311517a0}

\bibitem[{{Bolton} {et~al.}(2006{\natexlab{a}}){Bolton}, {Burles}, {Koopmans}, {Treu}, \& {Moustakas}}]{2006ApJ...638..703B}
{Bolton}, A.~S., {Burles}, S., {Koopmans}, L. V.~E., {Treu}, T., \& {Moustakas}, L.~A. 2006{\natexlab{a}}, \apj, 638, 703, \dodoi{10.1086/498884}

\bibitem[{{Bolton} {et~al.}(2006{\natexlab{b}}){Bolton}, {Burles}, {Koopmans}, {Treu}, \& {Moustakas}}]{Bolton:Burles:2006}
---. 2006{\natexlab{b}}, \apj, 638, 703, \dodoi{10.1086/498884}

\bibitem[{{Bolton} {et~al.}(2012){Bolton}, {Brownstein}, {Kochanek}, {Shu}, {Schlegel}, {Eisenstein}, {Wake}, {Connolly}, {Maraston}, {Arneson}, \& {Weaver}}]{Bolton:Brownstein:2012}
{Bolton}, A.~S., {Brownstein}, J.~R., {Kochanek}, C.~S., {et~al.} 2012, \apj, 757, 82, \dodoi{10.1088/0004-637X/757/1/82}

\bibitem[{{Brodie} {et~al.}(2014){Brodie}, {Romanowsky}, {Strader}, {Forbes}, {Foster}, {Jennings}, {Pastorello}, {Pota}, {Usher}, {Blom}, {Kader}, {Roediger}, {Spitler}, {Villaume}, {Arnold}, {Kartha}, \& {Woodley}}]{Brodie:Romanowsky:2014}
{Brodie}, J.~P., {Romanowsky}, A.~J., {Strader}, J., {et~al.} 2014, \apj, 796, 52, \dodoi{10.1088/0004-637X/796/1/52}

\bibitem[{{Brownstein} {et~al.}(2012){Brownstein}, {Bolton}, {Schlegel}, {Eisenstein}, {Kochanek}, {Connolly}, {Maraston}, {Pandey}, {Seitz}, {Wake}, {Wood-Vasey}, {Brinkmann}, {Schneider}, \& {Weaver}}]{Brownstein:Bolton:2012}
{Brownstein}, J.~R., {Bolton}, A.~S., {Schlegel}, D.~J., {et~al.} 2012, \apj, 744, 41, \dodoi{10.1088/0004-637X/744/1/41}

\bibitem[{{Bundy} {et~al.}(2015){Bundy}, {Bershady}, {Law}, {Yan}, {Drory}, {MacDonald}, {Wake}, {Cherinka}, {S{\'a}nchez-Gallego}, {Weijmans}, {Thomas}, {Tremonti}, {Masters}, {Coccato}, {Diamond-Stanic}, {Arag{\'o}n-Salamanca}, {Avila-Reese}, {Badenes}, {Falc{\'o}n-Barroso}, {Belfiore}, {Bizyaev}, {Blanc}, {Bland-Hawthorn}, {Blanton}, {Brownstein}, {Byler}, {Cappellari}, {Conroy}, {Dutton}, {Emsellem}, {Etherington}, {Frinchaboy}, {Fu}, {Gunn}, {Harding}, {Johnston}, {Kauffmann}, {Kinemuchi}, {Klaene}, {Knapen}, {Leauthaud}, {Li}, {Lin}, {Maiolino}, {Malanushenko}, {Malanushenko}, {Mao}, {Maraston}, {McDermid}, {Merrifield}, {Nichol}, {Oravetz}, {Pan}, {Parejko}, {Sanchez}, {Schlegel}, {Simmons}, {Steele}, {Steinmetz}, {Thanjavur}, {Thompson}, {Tinker}, {van den Bosch}, {Westfall}, {Wilkinson}, {Wright}, {Xiao}, \& {Zhang}}]{Bundy:Bershady:2015}
{Bundy}, K., {Bershady}, M.~A., {Law}, D.~R., {et~al.} 2015, \apj, 798, 7, \dodoi{10.1088/0004-637X/798/1/7}

\bibitem[{{Cabanac} {et~al.}(2007){Cabanac}, {Alard}, {Dantel-Fort}, {Fort}, {Gavazzi}, {Gomez}, {Kneib}, {Le F{\`e}vre}, {Mellier}, {Pello}, {Soucail}, {Sygnet}, \& {Valls-Gabaud}}]{Cabanac:Alard:2007}
{Cabanac}, R.~A., {Alard}, C., {Dantel-Fort}, M., {et~al.} 2007, \aap, 461, 813, \dodoi{10.1051/0004-6361:20065810}

\bibitem[{{Cappellari}(2012)}]{Cappellari:ppxf:2012}
{Cappellari}, M. 2012, {pPXF: Penalized Pixel-Fitting stellar kinematics extraction}, Astrophysics Source Code Library, record ascl:1210.002.
\newblock \doeprint{1210.002}

\bibitem[{{Cappellari}(2016)}]{Cappellari:2016}
---. 2016, \araa, 54, 597, \dodoi{10.1146/annurev-astro-082214-122432}

\bibitem[{{Cappellari}(2023)}]{Cappellari:2023}
---. 2023, \mnras, 526, 3273, \dodoi{10.1093/mnras/stad2597}

\bibitem[{{Cappellari} {et~al.}(2007){Cappellari}, {Emsellem}, {Bacon}, {Bureau}, {Davies}, {de Zeeuw}, {Falc{\'o}n-Barroso}, {Krajnovi{\'c}}, {Kuntschner}, {McDermid}, {Peletier}, {Sarzi}, {van den Bosch}, \& {van de Ven}}]{Cappellari:Emsellem:2007}
{Cappellari}, M., {Emsellem}, E., {Bacon}, R., {et~al.} 2007, \mnras, 379, 418, \dodoi{10.1111/j.1365-2966.2007.11963.x}

\bibitem[{{Chen} {et~al.}(2016){Chen}, {Hwang}, \& {Ko}}]{Chen:Hwang:2016}
{Chen}, C.-Y., {Hwang}, C.-Y., \& {Ko}, C.-M. 2016, \apj, 830, 123, \dodoi{10.3847/0004-637X/830/2/123}

\bibitem[{{Ciambur}(2016)}]{Ciambur:Profiler:2016}
{Ciambur}, B.~C. 2016, \pasa, 33, e062, \dodoi{10.1017/pasa.2016.60}

\bibitem[{{Ciotti} \& {Bertin}(1999)}]{Ciotti:Bertin:1999}
{Ciotti}, L., \& {Bertin}, G. 1999, \aap, 352, 447, \dodoi{10.48550/arXiv.astro-ph/9911078}

\bibitem[{{Coccato} {et~al.}(2009){Coccato}, {Gerhard}, {Arnaboldi}, {Das}, {Douglas}, {Kuijken}, {Merrifield}, {Napolitano}, {Noordermeer}, {Romanowsky}, {Capaccioli}, {Cortesi}, {De Lorenzi}, \& {Freeman}}]{Coccato:Gerhard:2009}
{Coccato}, L., {Gerhard}, O., {Arnaboldi}, M., {et~al.} 2009, \mnras, 394, 1249, \dodoi{10.1111/j.1365-2966.2009.14417.x}

\bibitem[{{Derkenne} {et~al.}(2021){Derkenne}, {McDermid}, {Poci}, {Remus}, {J{\o}rgensen}, \& {Emsellem}}]{Derkenne:McDermid:2021}
{Derkenne}, C., {McDermid}, R.~M., {Poci}, A., {et~al.} 2021, \mnras, 506, 3691, \dodoi{10.1093/mnras/stab1996}

\bibitem[{{Derkenne} {et~al.}(2023){Derkenne}, {McDermid}, {Poci}, {Mendel}, {D'Eugenio}, {Jeon}, {Remus}, {Bellstedt}, {Battisti}, {Bland-Hawthorn}, {Ferr{\'e}-Mateu}, {Foster}, {Harborne}, {Lagos}, {Peng}, {Sharda}, {Sharma}, {Sweet}, {Tran}, {Valenzuela}, {Vaughan}, {Wisnioski}, \& {Yi}}]{Derkenne:McDermid:Poci:2023}
---. 2023, \mnras, 522, 3602, \dodoi{10.1093/mnras/stad1079}

\bibitem[{{Dunkley} {et~al.}(2005){Dunkley}, {Bucher}, {Ferreira}, {Moodley}, \& {Skordis}}]{Dunkley:Bucher:2005}
{Dunkley}, J., {Bucher}, M., {Ferreira}, P.~G., {Moodley}, K., \& {Skordis}, C. 2005, \mnras, 356, 925, \dodoi{10.1111/j.1365-2966.2004.08464.x}

\bibitem[{{Dutton} \& {Treu}(2014)}]{Dutton:Treu:2014}
{Dutton}, A.~A., \& {Treu}, T. 2014, \mnras, 438, 3594, \dodoi{10.1093/mnras/stt2489}

\bibitem[{{Dutton} {et~al.}(2011){Dutton}, {Brewer}, {Marshall}, {Auger}, {Treu}, {Koo}, {Bolton}, {Holden}, \& {Koopmans}}]{Dutton:Brewer:2011}
{Dutton}, A.~A., {Brewer}, B.~J., {Marshall}, P.~J., {et~al.} 2011, \mnras, 417, 1621, \dodoi{10.1111/j.1365-2966.2011.18706.x}

\bibitem[{{Eisenstein} {et~al.}(2011){Eisenstein}, {Weinberg}, {Agol}, {Aihara}, {Allende Prieto}, {Anderson}, {Arns}, {Aubourg}, {Bailey}, {Balbinot}, {Barkhouser}, {Beers}, {Berlind}, {Bickerton}, {Bizyaev}, {Blanton}, {Bochanski}, {Bolton}, {Bosman}, {Bovy}, {Brandt}, {Breslauer}, {Brewington}, {Brinkmann}, {Brown}, {Brownstein}, {Burger}, {Busca}, {Campbell}, {Cargile}, {Carithers}, {Carlberg}, {Carr}, {Chang}, {Chen}, {Chiappini}, {Comparat}, {Connolly}, {Cortes}, {Croft}, {Cunha}, {da Costa}, {Davenport}, {Dawson}, {De Lee}, {Porto de Mello}, {de Simoni}, {Dean}, {Dhital}, {Ealet}, {Ebelke}, {Edmondson}, {Eiting}, {Escoffier}, {Esposito}, {Evans}, {Fan}, {Femen{\'\i}a Castell{\'a}}, {Dutra Ferreira}, {Fitzgerald}, {Fleming}, {Font-Ribera}, {Ford}, {Frinchaboy}, {Garc{\'\i}a P{\'e}rez}, {Gaudi}, {Ge}, {Ghezzi}, {Gillespie}, {Gilmore}, {Girardi}, {Gott}, {Gould}, {Grebel}, {Gunn}, {Hamilton}, {Harding}, {Harris}, {Hawley}, {Hearty}, {Hennawi}, {Gonz{\'a}lez Hern{\'a}ndez}, {Ho}, {Hogg}, {Holtzman},
  {Honscheid}, {Inada}, {Ivans}, {Jiang}, {Jiang}, {Johnson}, {Jordan}, {Jordan}, {Kauffmann}, {Kazin}, {Kirkby}, {Klaene}, {Knapp}, {Kneib}, {Kochanek}, {Koesterke}, {Kollmeier}, {Kron}, {Lampeitl}, {Lang}, {Lawler}, {Le Goff}, {Lee}, {Lee}, {Leisenring}, {Lin}, {Liu}, {Long}, {Loomis}, {Lucatello}, {Lundgren}, {Lupton}, {Ma}, {Ma}, {MacDonald}, {Mack}, {Mahadevan}, {Maia}, {Majewski}, {Makler}, {Malanushenko}, {Malanushenko}, {Mandelbaum}, {Maraston}, {Margala}, {Maseman}, {Masters}, {McBride}, {McDonald}, {McGreer}, {McMahon}, {Mena Requejo}, {M{\'e}nard}, {Miralda-Escud{\'e}}, {Morrison}, {Mullally}, {Muna}, {Murayama}, {Myers}, {Naugle}, {Neto}, {Nguyen}, {Nichol}, {Nidever}, {O'Connell}, {Ogando}, {Olmstead}, {Oravetz}, {Padmanabhan}, {Paegert}, {Palanque-Delabrouille}, {Pan}, {Pandey}, {Parejko}, {P{\^a}ris}, {Pellegrini}, {Pepper}, {Percival}, {Petitjean}, {Pfaffenberger}, {Pforr}, {Phleps}, {Pichon}, {Pieri}, {Prada}, {Price-Whelan}, {Raddick}, {Ramos}, {Reid}, {Reyle}, {Rich}, {Richards}, {Rieke},
  {Rieke}, {Rix}, {Robin}, {Rocha-Pinto}, {Rockosi}, {Roe}, {Rollinde}, {Ross}, {Ross}, {Rossetto}, {S{\'a}nchez}, {Santiago}, {Sayres}, {Schiavon}, {Schlegel}, {Schlesinger}, {Schmidt}, {Schneider}, {Sellgren}, {Shelden}, {Sheldon}, {Shetrone}, {Shu}, {Silverman}, {Simmerer}, {Simmons}, {Sivarani}, {Skrutskie}, {Slosar}, {Smee}, {Smith}, {Snedden}, {Stassun}, {Steele}, {Steinmetz}, {Stockett}, {Stollberg}, {Strauss}, {Szalay}, {Tanaka}, {Thakar}, {Thomas}, {Tinker}, {Tofflemire}, {Tojeiro}, {Tremonti}, {Vargas Maga{\~n}a}, {Verde}, {Vogt}, {Wake}, {Wan}, {Wang}, {Weaver}, {White}, {White}, {Wilson}, {Wisniewski}, {Wood-Vasey}, {Yanny}, {Yasuda}, {Y{\`e}che}, {York}, {Young}, {Zasowski}, {Zehavi}, \& {Zhao}}]{Eisenstein:Weinberg:2011}
{Eisenstein}, D.~J., {Weinberg}, D.~H., {Agol}, E., {et~al.} 2011, \aj, 142, 72, \dodoi{10.1088/0004-6256/142/3/72}

\bibitem[{{Ertl} {et~al.}(2023){Ertl}, {Schuldt}, {Suyu}, {Schmidt}, {Treu}, {Birrer}, {Shajib}, \& {Sluse}}]{Ertl:Schuldt:2023}
{Ertl}, S., {Schuldt}, S., {Suyu}, S.~H., {et~al.} 2023, \aap, 672, A2, \dodoi{10.1051/0004-6361/202244909}

\bibitem[{{Etherington} {et~al.}(2023){Etherington}, {Nightingale}, {Massey}, {Robertson}, {Cao}, {Amvrosiadis}, {Cole}, {Frenk}, {He}, {Lagattuta}, {Lange}, \& {Li}}]{Etherington:Nightingale:2023}
{Etherington}, A., {Nightingale}, J.~W., {Massey}, R., {et~al.} 2023, \mnras, 521, 6005, \dodoi{10.1093/mnras/stad582}

\bibitem[{{Falco} {et~al.}(1985){Falco}, {Gorenstein}, \& {Shapiro}}]{Falco:Gorenstein:1985}
{Falco}, E.~E., {Gorenstein}, M.~V., \& {Shapiro}, I.~I. 1985, \apjl, 289, L1, \dodoi{10.1086/184422}

\bibitem[{{Filipp} {et~al.}(2023){Filipp}, {Shu}, {Pakmor}, {Suyu}, \& {Huang}}]{Filipp:Shu:2023}
{Filipp}, A., {Shu}, Y., {Pakmor}, R., {Suyu}, S.~H., \& {Huang}, X. 2023, arXiv e-prints, arXiv:2307.15044, \dodoi{10.48550/arXiv.2307.15044}

\bibitem[{{Foreman-Mackey} {et~al.}(2013){Foreman-Mackey}, {Hogg}, {Lang}, \& {Goodman}}]{Foreman-Mackey:Hogg:2013}
{Foreman-Mackey}, D., {Hogg}, D.~W., {Lang}, D., \& {Goodman}, J. 2013, \pasp, 125, 306, \dodoi{10.1086/670067}

\bibitem[{{Forrest} {et~al.}(2022){Forrest}, {Wilson}, {Muzzin}, {Marchesini}, {Cooper}, {Marsan}, {Annunziatella}, {McConachie}, {Zaidi}, {Gomez}, {Urbano Stawinski}, {Chang}, {de Lucia}, {La Barbera}, {Lubin}, {Nantais}, {Pe{\~n}a}, {Saracco}, {Surace}, \& {Stefanon}}]{Forrest:Wilson:2022}
{Forrest}, B., {Wilson}, G., {Muzzin}, A., {et~al.} 2022, \apj, 938, 109, \dodoi{10.3847/1538-4357/ac8747}

\bibitem[{{Foster} {et~al.}(2021){Foster}, {Mendel}, {Lagos}, {Wisnioski}, {Yuan}, {D'Eugenio}, {Barone}, {Harborne}, {Vaughan}, {Schulze}, {Remus}, {Gupta}, {Collacchioni}, {Khim}, {Taylor}, {Bassett}, {Croom}, {McDermid}, {Poci}, {Battisti}, {Bland-Hawthorn}, {Bellstedt}, {Colless}, {Davies}, {Derkenne}, {Driver}, {Ferr{\'e}-Mateu}, {Fisher}, {Gjergo}, {Johnston}, {Khalid}, {Kobayashi}, {Oh}, {Peng}, {Robotham}, {Sharda}, {Sweet}, {Taylor}, {Tran}, {Trayford}, {van de Sande}, {Yi}, \& {Zanisi}}]{Foster:Mendel:2021}
{Foster}, C., {Mendel}, J.~T., {Lagos}, C.~D.~P., {et~al.} 2021, \pasa, 38, e031, \dodoi{10.1017/pasa.2021.25}

\bibitem[{{Guo} \& {White}(2008)}]{Guo:White:2008}
{Guo}, Q., \& {White}, S.~D.~M. 2008, \mnras, 384, 2, \dodoi{10.1111/j.1365-2966.2007.12619.x}

\bibitem[{{Halkola} {et~al.}(2008){Halkola}, {Hildebrandt}, {Schrabback}, {Lombardi}, {Brada{\v{c}}}, {Erben}, {Schneider}, \& {Wuttke}}]{Halkola:2008}
{Halkola}, A., {Hildebrandt}, H., {Schrabback}, T., {et~al.} 2008, \aap, 481, 65, \dodoi{10.1051/0004-6361:20078877}

\bibitem[{{Halkola} {et~al.}(2006){Halkola}, {Seitz}, \& {Pannella}}]{Halkola:2006}
{Halkola}, A., {Seitz}, S., \& {Pannella}, M. 2006, \mnras, 372, 1425, \dodoi{10.1111/j.1365-2966.2006.10948.x}

\bibitem[{{Harris} {et~al.}(2020){Harris}, {Millman}, {van der Walt}, {Gommers}, {Virtanen}, {Cournapeau}, {Wieser}, {Taylor}, {Berg}, {Smith}, {Kern}, {Picus}, {Hoyer}, {van Kerkwijk}, {Brett}, {Haldane}, {del R{\'\i}o}, {Wiebe}, {Peterson}, {G{\'e}rard-Marchant}, {Sheppard}, {Reddy}, {Weckesser}, {Abbasi}, {Gohlke}, \& {Oliphant}}]{NumPy:2020}
{Harris}, C.~R., {Millman}, K.~J., {van der Walt}, S.~J., {et~al.} 2020, \nat, 585, 357, \dodoi{10.1038/s41586-020-2649-2}

\bibitem[{{Humphrey} \& {Buote}(2010)}]{Humphrey:Buote:2010}
{Humphrey}, P.~J., \& {Buote}, D.~A. 2010, \mnras, 403, 2143, \dodoi{10.1111/j.1365-2966.2010.16257.x}

\bibitem[{Hunter(2007)}]{matplotlib:2007}
Hunter, J.~D. 2007, Computing in science \& engineering, 9, 90

\bibitem[{{Kirkpatrick} {et~al.}(1983){Kirkpatrick}, {Gelatt}, \& {Vecchi}}]{Kirkpatrick:Gelatt:1983}
{Kirkpatrick}, S., {Gelatt}, C.~D., \& {Vecchi}, M.~P. 1983, Science, 220, 671, \dodoi{10.1126/science.220.4598.671}

\bibitem[{{Kochanek}(2020)}]{Kochanek:2020}
{Kochanek}, C.~S. 2020, \mnras, 493, 1725, \dodoi{10.1093/mnras/staa344}

\bibitem[{{Koopmans} \& {Treu}(2004)}]{Koopmans:Treu:2004}
{Koopmans}, L.~V.~E., \& {Treu}, T. 2004, in Astrophysics and Space Science Library, Vol. 301, Astrophysics and Space Science Library, ed. M.~{Plionis}, 23, \dodoi{10.1007/0-306-48570-2_4}

\bibitem[{{Koopmans} {et~al.}(2006){Koopmans}, {Treu}, {Bolton}, {Burles}, \& {Moustakas}}]{Koopmans:Treu:Bolton:2006}
{Koopmans}, L. V.~E., {Treu}, T., {Bolton}, A.~S., {Burles}, S., \& {Moustakas}, L.~A. 2006, \apj, 649, 599, \dodoi{10.1086/505696}

\bibitem[{{Koopmans} {et~al.}(2009){Koopmans}, {Bolton}, {Treu}, {Czoske}, {Auger}, {Barnab{\`e}}, {Vegetti}, {Gavazzi}, {Moustakas}, \& {Burles}}]{Koopmans:Bolton:2009}
{Koopmans}, L.~V.~E., {Bolton}, A., {Treu}, T., {et~al.} 2009, \apjl, 703, L51, \dodoi{10.1088/0004-637X/703/1/L51}

\bibitem[{{Krist} {et~al.}(2011){Krist}, {Hook}, \& {Stoehr}}]{Krist:2011}
{Krist}, J.~E., {Hook}, R.~N., \& {Stoehr}, F. 2011, in Society of Photo-Optical Instrumentation Engineers (SPIE) Conference Series, Vol. 8127, Optical Modeling and Performance Predictions V, ed. M.~A. {Kahan}, 81270J, \dodoi{10.1117/12.892762}

\bibitem[{{Kronawitter} {et~al.}(2000){Kronawitter}, {Saglia}, {Gerhard}, \& {Bender}}]{Kronawitter:Saglia:2000}
{Kronawitter}, A., {Saglia}, R.~P., {Gerhard}, O., \& {Bender}, R. 2000, \aaps, 144, 53, \dodoi{10.1051/aas:2000199}

\bibitem[{{Li} {et~al.}(2018{\natexlab{a}}){Li}, {Shu}, \& {Wang}}]{Li:Shu:Wang:2018}
{Li}, R., {Shu}, Y., \& {Wang}, J. 2018{\natexlab{a}}, \mnras, 480, 431, \dodoi{10.1093/mnras/sty1813}

\bibitem[{{Li} {et~al.}(2018{\natexlab{b}}){Li}, {Wang}, {Shu}, \& {Xu}}]{Li:Wang:2018}
{Li}, R., {Wang}, J., {Shu}, Y., \& {Xu}, Z. 2018{\natexlab{b}}, \apj, 855, 64, \dodoi{10.3847/1538-4357/aaab50}

\bibitem[{{Li} {et~al.}(2019){Li}, {Li}, {Shao}, {Lu}, {Zhu}, {Wang}, {Gao}, {Mao}, {Dutton}, {Ge}, {Wang}, {Leauthaud}, {Zheng}, {Bundy}, \& {Brownstein}}]{Li:Li:Shao:2019}
{Li}, R., {Li}, H., {Shao}, S., {et~al.} 2019, \mnras, 490, 2124, \dodoi{10.1093/mnras/stz2565}

\bibitem[{{McKerns} {et~al.}(2012){McKerns}, {Strand}, {Sullivan}, {Fang}, \& {Aivazis}}]{Multiprocess:McKerns:Strand:2012}
{McKerns}, M.~M., {Strand}, L., {Sullivan}, T., {Fang}, A., \& {Aivazis}, M. A.~G. 2012, arXiv e-prints, arXiv:1202.1056, \dodoi{10.48550/arXiv.1202.1056}

\bibitem[{{Merritt}(1985)}]{Merritt:1985}
{Merritt}, D. 1985, \aj, 90, 1027, \dodoi{10.1086/113810}

\bibitem[{{Mihos} \& {Hernquist}(1994)}]{Mihos:Hernquist:1994}
{Mihos}, J.~C., \& {Hernquist}, L. 1994, \apjl, 425, L13, \dodoi{10.1086/187299}

\bibitem[{{Mukherjee} {et~al.}(2021){Mukherjee}, {Koopmans}, {Metcalf}, {Tortora}, {Schaller}, {Schaye}, {Vernardos}, \& {Bellagamba}}]{Mukherjee:Koopmans:2021}
{Mukherjee}, S., {Koopmans}, L. V.~E., {Metcalf}, R.~B., {et~al.} 2021, \mnras, 504, 3455, \dodoi{10.1093/mnras/stab693}

\bibitem[{{Naab} {et~al.}(2007){Naab}, {Johansson}, {Ostriker}, \& {Efstathiou}}]{Naab:Johansson:2007}
{Naab}, T., {Johansson}, P.~H., {Ostriker}, J.~P., \& {Efstathiou}, G. 2007, \apj, 658, 710, \dodoi{10.1086/510841}

\bibitem[{{Newman} {et~al.}(2015){Newman}, {Ellis}, \& {Treu}}]{Newman:Ellis:2015}
{Newman}, A.~B., {Ellis}, R.~S., \& {Treu}, T. 2015, \apj, 814, 26, \dodoi{10.1088/0004-637X/814/1/26}

\bibitem[{{Oser} {et~al.}(2010){Oser}, {Ostriker}, {Naab}, {Johansson}, \& {Burkert}}]{Oser:Ostriker:2010}
{Oser}, L., {Ostriker}, J.~P., {Naab}, T., {Johansson}, P.~H., \& {Burkert}, A. 2010, \apj, 725, 2312, \dodoi{10.1088/0004-637X/725/2/2312}

\bibitem[{{Osipkov}(1979)}]{Osipkov:1979}
{Osipkov}, L.~P. 1979, Pisma v Astronomicheskii Zhurnal, 5, 77

\bibitem[{{Peirani} {et~al.}(2019){Peirani}, {Sonnenfeld}, {Gavazzi}, {Oguri}, {Dubois}, {Silk}, {Pichon}, {Devriendt}, \& {Kaviraj}}]{Peirani:Sonnenfeld:Gavazzi:2019}
{Peirani}, S., {Sonnenfeld}, A., {Gavazzi}, R., {et~al.} 2019, \mnras, 483, 4615, \dodoi{10.1093/mnras/sty3475}

\bibitem[{{Poci} {et~al.}(2017){Poci}, {Cappellari}, \& {McDermid}}]{Poci:Cappellari:2017}
{Poci}, A., {Cappellari}, M., \& {McDermid}, R.~M. 2017, \mnras, 467, 1397, \dodoi{10.1093/mnras/stx101}

\bibitem[{{Remus} {et~al.}(2017){Remus}, {Dolag}, {Naab}, {Burkert}, {Hirschmann}, {Hoffmann}, \& {Johansson}}]{Remus:Dolag:Naab:2017}
{Remus}, R.-S., {Dolag}, K., {Naab}, T., {et~al.} 2017, \mnras, 464, 3742, \dodoi{10.1093/mnras/stw2594}

\bibitem[{{Ruff} {et~al.}(2011){Ruff}, {Gavazzi}, {Marshall}, {Treu}, {Auger}, \& {Brault}}]{Ruff:Gavazzi:2011}
{Ruff}, A.~J., {Gavazzi}, R., {Marshall}, P.~J., {et~al.} 2011, \apj, 727, 96, \dodoi{10.1088/0004-637X/727/2/96}

\bibitem[{{Saglia} {et~al.}(2000){Saglia}, {Kronawitter}, {Gerhard}, \& {Bender}}]{Saglia:Kronawitter:2000}
{Saglia}, R.~P., {Kronawitter}, A., {Gerhard}, O., \& {Bender}, R. 2000, \aj, 119, 153, \dodoi{10.1086/301153}

\bibitem[{{S{\'a}nchez-Bl{\'a}zquez} {et~al.}(2006){S{\'a}nchez-Bl{\'a}zquez}, {Peletier}, {Jim{\'e}nez-Vicente}, {Cardiel}, {Cenarro}, {Falc{\'o}n-Barroso}, {Gorgas}, {Selam}, \& {Vazdekis}}]{MILES:Sanchez-Blazquez:2006}
{S{\'a}nchez-Bl{\'a}zquez}, P., {Peletier}, R.~F., {Jim{\'e}nez-Vicente}, J., {et~al.} 2006, \mnras, 371, 703, \dodoi{10.1111/j.1365-2966.2006.10699.x}

\bibitem[{{Schneider} {et~al.}(1992){Schneider}, {Ehlers}, \& {Falco}}]{Schneider:Ehlers:Falco:1992}
{Schneider}, P., {Ehlers}, J., \& {Falco}, E.~E. 1992, {Gravitational Lenses}, \dodoi{10.1007/978-3-662-03758-4}

\bibitem[{{Schneider} \& {Sluse}(2013)}]{Schneider:Sluse:2013}
{Schneider}, P., \& {Sluse}, D. 2013, \aap, 559, A37, \dodoi{10.1051/0004-6361/201321882}

\bibitem[{{S{\'e}rsic}(1963)}]{Sersic:1963}
{S{\'e}rsic}, J.~L. 1963, Boletin de la Asociacion Argentina de Astronomia La Plata Argentina, 6, 41

\bibitem[{{Shajib} {et~al.}(2021){Shajib}, {Treu}, {Birrer}, \& {Sonnenfeld}}]{Shajib:Treu:Birrer:2021}
{Shajib}, A.~J., {Treu}, T., {Birrer}, S., \& {Sonnenfeld}, A. 2021, \mnras, 503, 2380, \dodoi{10.1093/mnras/stab536}

\bibitem[{{Shajib} {et~al.}(2022{\natexlab{a}}){Shajib}, {Vernardos}, {Collett}, {Motta}, {Sluse}, {Williams}, {Saha}, {Birrer}, {Spiniello}, \& {Treu}}]{Shajib:Vernardos:Collett:2022}
{Shajib}, A.~J., {Vernardos}, G., {Collett}, T.~E., {et~al.} 2022{\natexlab{a}}, arXiv e-prints, arXiv:2210.10790, \dodoi{10.48550/arXiv.2210.10790}

\bibitem[{{Shajib} {et~al.}(2022{\natexlab{b}}){Shajib}, {Glazebrook}, {Barone}, {Lewis}, {Jones}, {Tran}, {Buckley-Geer}, {Collett}, {Frieman}, \& {Jacobs}}]{Shajib:Glazebrook:Barone:2022}
{Shajib}, A.~J., {Glazebrook}, K., {Barone}, T., {et~al.} 2022{\natexlab{b}}, \apj, 938, 141, \dodoi{10.3847/1538-4357/ac927b}

\bibitem[{{Shajib} {et~al.}(2023){Shajib}, {Mozumdar}, {Chen}, {Treu}, {Cappellari}, {Knabel}, {Suyu}, {Bennert}, {Frieman}, {Sluse}, {Birrer}, {Courbin}, {Fassnacht}, {Villafa{\~n}a}, \& {Williams}}]{Shajib:Mozumdar:2023}
{Shajib}, A.~J., {Mozumdar}, P., {Chen}, G. C.~F., {et~al.} 2023, \aap, 673, A9, \dodoi{10.1051/0004-6361/202345878}

\bibitem[{{Sheinis} {et~al.}(2002){Sheinis}, {Bolte}, {Epps}, {Kibrick}, {Miller}, {Radovan}, {Bigelow}, \& {Sutin}}]{ESI:Sheinis:2002}
{Sheinis}, A.~I., {Bolte}, M., {Epps}, H.~W., {et~al.} 2002, \pasp, 114, 851, \dodoi{10.1086/341706}

\bibitem[{{Shu} {et~al.}(2016){Shu}, {Bolton}, {Mao}, {Kochanek}, {P{\'e}rez-Fournon}, {Oguri}, {Montero-Dorta}, {Cornachione}, {Marques-Chaves}, {Zheng}, {Brownstein}, \& {M{\'e}nard}}]{Shu:Bolton:2016}
{Shu}, Y., {Bolton}, A.~S., {Mao}, S., {et~al.} 2016, \apj, 833, 264, \dodoi{10.3847/1538-4357/833/2/264}

\bibitem[{{Somerville} \& {Dav{\'e}}(2015)}]{Somerville:Dave:2015}
{Somerville}, R.~S., \& {Dav{\'e}}, R. 2015, \araa, 53, 51, \dodoi{10.1146/annurev-astro-082812-140951}

\bibitem[{{Sonnenfeld} {et~al.}(2013{\natexlab{a}}){Sonnenfeld}, {Gavazzi}, {Suyu}, {Treu}, \& {Marshall}}]{Sonnenfeld:Gavazzi:2013}
{Sonnenfeld}, A., {Gavazzi}, R., {Suyu}, S.~H., {Treu}, T., \& {Marshall}, P.~J. 2013{\natexlab{a}}, \apj, 777, 97, \dodoi{10.1088/0004-637X/777/2/97}

\bibitem[{{Sonnenfeld} {et~al.}(2014){Sonnenfeld}, {Nipoti}, \& {Treu}}]{Sonnenfeld:Nipoti:2014}
{Sonnenfeld}, A., {Nipoti}, C., \& {Treu}, T. 2014, \apj, 786, 89, \dodoi{10.1088/0004-637X/786/2/89}

\bibitem[{{Sonnenfeld} {et~al.}(2013{\natexlab{b}}){Sonnenfeld}, {Treu}, {Gavazzi}, {Suyu}, {Marshall}, {Auger}, \& {Nipoti}}]{Sonnenfeld:Treu:Gavazzi:2013}
{Sonnenfeld}, A., {Treu}, T., {Gavazzi}, R., {et~al.} 2013{\natexlab{b}}, \apj, 777, 98, \dodoi{10.1088/0004-637X/777/2/98}

\bibitem[{{Suyu} \& {Halkola}(2010)}]{Suyu:Halkola:2010}
{Suyu}, S.~H., \& {Halkola}, A. 2010, \aap, 524, A94, \dodoi{10.1051/0004-6361/201015481}

\bibitem[{{Suyu} {et~al.}(2006){Suyu}, {Marshall}, {Hobson}, \& {Blandford}}]{Suyu:Marshall:Hobson:2006}
{Suyu}, S.~H., {Marshall}, P.~J., {Hobson}, M.~P., \& {Blandford}, R.~D. 2006, \mnras, 371, 983, \dodoi{10.1111/j.1365-2966.2006.10733.x}

\bibitem[{{Suyu} {et~al.}(2012){Suyu}, {Hensel}, {McKean}, {Fassnacht}, {Treu}, {Halkola}, {Norbury}, {Jackson}, {Schneider}, {Thompson}, {Auger}, {Koopmans}, \& {Matthews}}]{Suyu:Hensel:McKean:2012}
{Suyu}, S.~H., {Hensel}, S.~W., {McKean}, J.~P., {et~al.} 2012, \apj, 750, 10, \dodoi{10.1088/0004-637X/750/1/10}

\bibitem[{{Suyu} {et~al.}(2013){Suyu}, {Auger}, {Hilbert}, {Marshall}, {Tewes}, {Treu}, {Fassnacht}, {Koopmans}, {Sluse}, {Blandford}, {Courbin}, \& {Meylan}}]{Suyu:Auger:Hilbert:2013}
{Suyu}, S.~H., {Auger}, M.~W., {Hilbert}, S., {et~al.} 2013, \apj, 766, 70, \dodoi{10.1088/0004-637X/766/2/70}

\bibitem[{{Tan} {et~al.}(2024){Tan}, {Shajib}, {Birrer}, {Sonnenfeld}, {Treu}, {Wells}, {Williams}, {Buckley-Geer}, {Drlica-Wagner}, \& {Frieman}}]{Tan:Shajib:2023}
{Tan}, C.~Y., {Shajib}, A.~J., {Birrer}, S., {et~al.} 2024, \mnras, \dodoi{10.1093/mnras/stae884}

\bibitem[{{Thomas} {et~al.}(2013){Thomas}, {Steele}, {Maraston}, {Johansson}, {Beifiori}, {Pforr}, {Str{\"o}mb{\"a}ck}, {Tremonti}, {Wake}, {Bizyaev}, {Bolton}, {Brewington}, {Brownstein}, {Comparat}, {Kneib}, {Malanushenko}, {Malanushenko}, {Oravetz}, {Pan}, {Parejko}, {Schneider}, {Shelden}, {Simmons}, {Snedden}, {Tanaka}, {Weaver}, \& {Yan}}]{Thomas:Steele:2013}
{Thomas}, D., {Steele}, O., {Maraston}, C., {et~al.} 2013, \mnras, 431, 1383, \dodoi{10.1093/mnras/stt261}

\bibitem[{{Thomas} {et~al.}(2011){Thomas}, {Saglia}, {Bender}, {Thomas}, {Gebhardt}, {Magorrian}, {Corsini}, {Wegner}, \& {Seitz}}]{Thomas:Saglia:2011}
{Thomas}, J., {Saglia}, R.~P., {Bender}, R., {et~al.} 2011, \mnras, 415, 545, \dodoi{10.1111/j.1365-2966.2011.18725.x}

\bibitem[{{Tran} {et~al.}(2022){Tran}, {Harshan}, {Glazebrook}, {Keerthi Vasan}, {Jones}, {Jacobs}, {Kacprzak}, {Barone}, {Collett}, {Gupta}, {Henderson}, {Kewley}, {Lopez}, {Nanayakkara}, {Sanders}, \& {Sweet}}]{Tran:Harshan:2022}
{Tran}, K.-V.~H., {Harshan}, A., {Glazebrook}, K., {et~al.} 2022, \aj, 164, 148, \dodoi{10.3847/1538-3881/ac7da2}

\bibitem[{{Treu}(2010)}]{Treu:2010}
{Treu}, T. 2010, \araa, 48, 87, \dodoi{10.1146/annurev-astro-081309-130924}

\bibitem[{{Treu} \& {Koopmans}(2002)}]{Treu:Koopmans:2002a}
{Treu}, T., \& {Koopmans}, L. V.~E. 2002, \apj, 575, 87, \dodoi{10.1086/341216}

\bibitem[{{Treu} \& {Koopmans}(2004)}]{Treu:Koopmans:2004}
---. 2004, \apj, 611, 739, \dodoi{10.1086/422245}

\bibitem[{{van de Sande} {et~al.}(2017){van de Sande}, {Bland-Hawthorn}, {Fogarty}, {Cortese}, {d'Eugenio}, {Croom}, {Scott}, {Allen}, {Brough}, {Bryant}, {Cecil}, {Colless}, {Couch}, {Davies}, {Elahi}, {Foster}, {Goldstein}, {Goodwin}, {Groves}, {Ho}, {Jeong}, {Jones}, {Konstantopoulos}, {Lawrence}, {Leslie}, {L{\'o}pez-S{\'a}nchez}, {McDermid}, {McElroy}, {Medling}, {Oh}, {Owers}, {Richards}, {Schaefer}, {Sharp}, {Sweet}, {Taranu}, {Tonini}, {Walcher}, \& {Yi}}]{vandeSande:Bland-Hawthorn:2017}
{van de Sande}, J., {Bland-Hawthorn}, J., {Fogarty}, L. M.~R., {et~al.} 2017, \apj, 835, 104, \dodoi{10.3847/1538-4357/835/1/104}

\bibitem[{{Vernet} {et~al.}(2011){Vernet}, {Dekker}, {D'Odorico}, {Kaper}, {Kjaergaard}, {Hammer}, {Randich}, {Zerbi}, {Groot}, {Hjorth}, {Guinouard}, {Navarro}, {Adolfse}, {Albers}, {Amans}, {Andersen}, {Andersen}, {Binetruy}, {Bristow}, {Castillo}, {Chemla}, {Christensen}, {Conconi}, {Conzelmann}, {Dam}, {de Caprio}, {de Ugarte Postigo}, {Delabre}, {di Marcantonio}, {Downing}, {Elswijk}, {Finger}, {Fischer}, {Flores}, {Fran{\c{c}}ois}, {Goldoni}, {Guglielmi}, {Haigron}, {Hanenburg}, {Hendriks}, {Horrobin}, {Horville}, {Jessen}, {Kerber}, {Kern}, {Kiekebusch}, {Kleszcz}, {Klougart}, {Kragt}, {Larsen}, {Lizon}, {Lucuix}, {Mainieri}, {Manuputy}, {Martayan}, {Mason}, {Mazzoleni}, {Michaelsen}, {Modigliani}, {Moehler}, {M{\o}ller}, {Norup S{\o}rensen}, {N{\o}rregaard}, {P{\'e}roux}, {Patat}, {Pena}, {Pragt}, {Reinero}, {Rigal}, {Riva}, {Roelfsema}, {Royer}, {Sacco}, {Santin}, {Schoenmaker}, {Spano}, {Sweers}, {Ter Horst}, {Tintori}, {Tromp}, {van Dael}, {van der Vliet}, {Venema}, {Vidali}, {Vinther}, {Vola},
  {Winters}, {Wistisen}, {Wulterkens}, \& {Zacchei}}]{Xshooter:Vernet:2011}
{Vernet}, J., {Dekker}, H., {D'Odorico}, S., {et~al.} 2011, \aap, 536, A105, \dodoi{10.1051/0004-6361/201117752}

\bibitem[{{Virtanen} {et~al.}(2020){Virtanen}, {Gommers}, {Oliphant}, {Haberland}, {Reddy}, {Cournapeau}, {Burovski}, {Peterson}, {Weckesser}, {Bright}, {van der Walt}, {Brett}, {Wilson}, {Millman}, {Mayorov}, {Nelson}, {Jones}, {Kern}, {Larson}, {Carey}, {Polat}, {Feng}, {Moore}, {VanderPlas}, {Laxalde}, {Perktold}, {Cimrman}, {Henriksen}, {Quintero}, {Harris}, {Archibald}, {Ribeiro}, {Pedregosa}, {van Mulbregt}, \& {SciPy 1. 0 Contributors}}]{SciPy:2020}
{Virtanen}, P., {Gommers}, R., {Oliphant}, T.~E., {et~al.} 2020, Nature Methods, 17, 261, \dodoi{10.1038/s41592-019-0686-2}

\bibitem[{{Wang} {et~al.}(2019){Wang}, {Vogelsberger}, {Xu}, {Shen}, {Mao}, {Barnes}, {Li}, {Marinacci}, {Torrey}, {Springel}, \& {Hernquist}}]{Wang:Vogelsberger:2019}
{Wang}, Y., {Vogelsberger}, M., {Xu}, D., {et~al.} 2019, \mnras, 490, 5722, \dodoi{10.1093/mnras/stz2907}

\bibitem[{{Wang} {et~al.}(2020){Wang}, {Vogelsberger}, {Xu}, {Mao}, {Springel}, {Li}, {Barnes}, {Hernquist}, {Pillepich}, {Marinacci}, {Pakmor}, {Weinberger}, \& {Torrey}}]{Wang:Vogelsberger:2020}
---. 2020, \mnras, 491, 5188, \dodoi{10.1093/mnras/stz3348}

\bibitem[{{Weijmans} {et~al.}(2008){Weijmans}, {Krajnovi{\'c}}, {van de Ven}, {Oosterloo}, {Morganti}, \& {de Zeeuw}}]{Weijmans:Krajnovic:2008}
{Weijmans}, A.-M., {Krajnovi{\'c}}, D., {van de Ven}, G., {et~al.} 2008, \mnras, 383, 1343, \dodoi{10.1111/j.1365-2966.2007.12680.x}

\bibitem[{{White} \& {Frenk}(1991)}]{White:Frenk:1991}
{White}, S. D.~M., \& {Frenk}, C.~S. 1991, \apj, 379, 52, \dodoi{10.1086/170483}

\bibitem[{{Xu} {et~al.}(2017){Xu}, {Springel}, {Sluse}, {Schneider}, {Sonnenfeld}, {Nelson}, {Vogelsberger}, \& {Hernquist}}]{Xu:Springel:Sluse:2017}
{Xu}, D., {Springel}, V., {Sluse}, D., {et~al.} 2017, \mnras, 469, 1824, \dodoi{10.1093/mnras/stx899}

\end{thebibliography}

\appendix
\renewcommand{\thefigure}{A\arabic{figure}}
\renewcommand{\thetable}{A\arabic{table}}
\setcounter{figure}{0}
\setcounter{table}{0}

\section{Lens Modeling Results}
\label{appendix: lens models}
Table \ref{table:lens parameters} provides the model parameters for the SPEMD density profile,  external shear component, and light profile parameters captured using two S\'ersic functions for the deflector galaxies in \agel \, lenses modeled here.  
Furthermore, we also provide single S\'ersic fit parameters for the lens light. However, single S\'ersic fit does not properly  capture the total galaxy light, especially the central light; hence, single S\'ersic fit parameter such as $R_{\rm eff}$ does not represent the actual half-light radius of the lens galaxy.
The modeling procedure can be found in Section \ref{Sec:modelling}.
 Galaxy apparent magnitude in the HST/F140W band (AB mag) obtained by integrating the flux captured by two S\'ersic components and the overall galaxy half-light radii $R_{\rm half, gal}$ along the semi-major axis for the deflector galaxies are already presented in Table \ref{table:redshift}. The circularized half-light radius, $R_{\rm half, gal, eq}$, obtained via $R_{\rm half, gal, eq}=R_{\rm half, gal} \sqrt{q_{\rm L}}$ \citep[see][]{Ciambur:Profiler:2016} is also provided in Table \ref{table:lens parameters}.

\begin{figure*}[h]
\begin{center}
\includegraphics[clip=true,trim= 10mm 0mm 10mm 05mm,width=  0.91\textwidth]{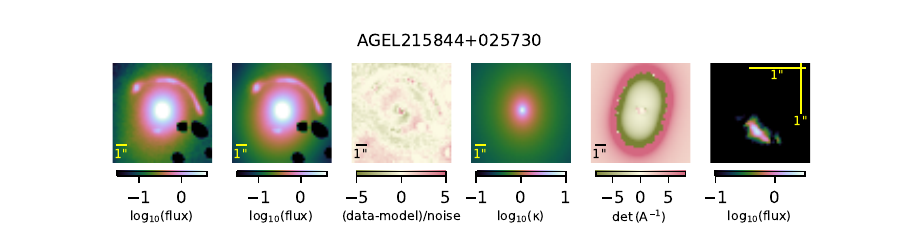}
\includegraphics[clip=true,trim= 10mm 00mm 10mm 05mm,width=  0.91\textwidth]{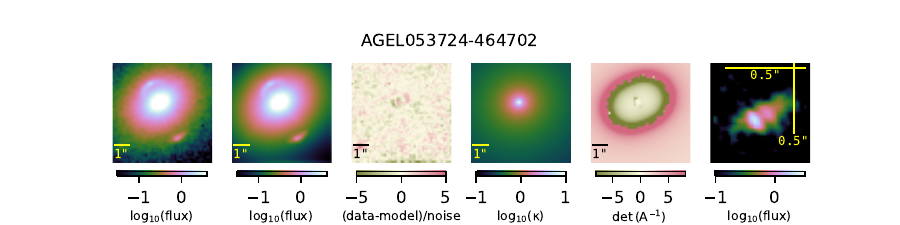}
\includegraphics[clip=true,trim= 10mm 00mm 10mm 05mm,width=  0.91\textwidth]{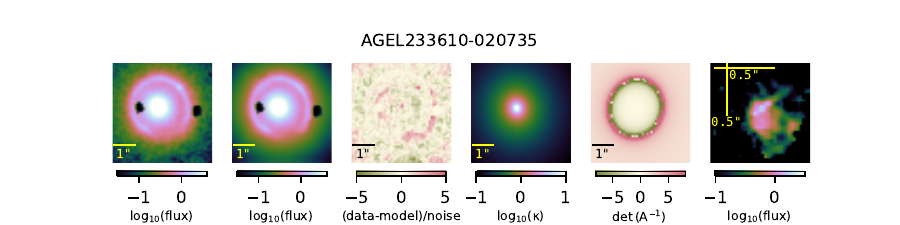}
\includegraphics[clip=true,trim= 10mm 00mm 10mm 05mm,width=  0.91\textwidth]{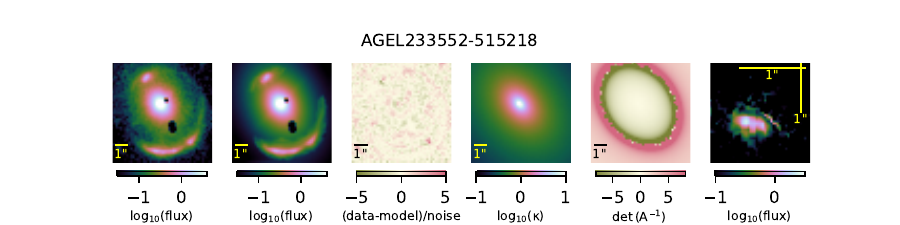}
\includegraphics[clip=true,trim= 10mm 00mm 10mm 05mm,width=  0.91\textwidth]{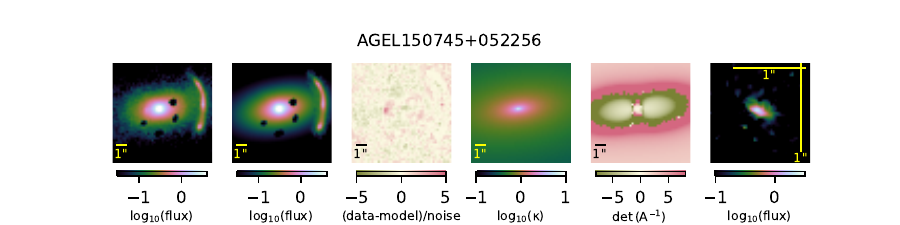}
\includegraphics[clip=true,trim= 10mm 00mm 10mm 05mm,width=  0.91\textwidth]{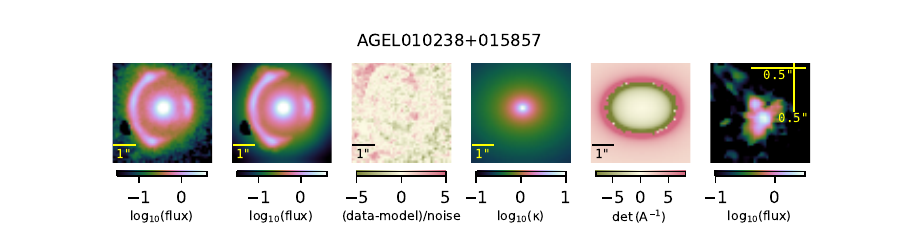}
\caption{Lens modeling results for remaining six \agel \, lenses modeled here using \glee. See sections \ref{Result:lens_models} and \ref{appendix: lens models} for details.}
\label{fig:models appendix}
\end{center}
\end{figure*}

Figure \ref{fig:models appendix} shows the most probable lens model of the remaining six lenses: \agel2158, \agel0537, \agel2336, \agel2335, \agel1507, and \agel0102. 
From left to right, each panel in Figure \ref{fig:models appendix} shows the observed HST  image, predicted model of the lens system, normalized residual, convergence map, magnification model, and the reconstructed source.
Here, flux is in the units of electron count per second. The first five panels are in the deflector plane. The right-most panel is in the source plane. 
The first five panels have the same grid size and pixel resolution; however, the source grid in the right-most panel has a higher resolution that depends on the observed image resolution and magnification caused by the lens. 
The angular sizes for the lens images and background source grid are provided in Table \ref{table:redshift}.

\begin{table}[h]
    \centering
    \begin{tabular}{|l|rrrrrrr|} \hline 
       \textbf{Parameter} &  \textbf{\agel2158}&  \textbf{\agel0537}&  \textbf{\agel2336}&  \textbf{\agel2335}&  \textbf{\agel1507}&  \textbf{\agel0142} &  \textbf{\agel0102}\\
        \hline
        \hline
          & \multicolumn{6}{l}{SPEMD} & \\
         \hline
           \bm{$\gamma^{\rm lens}$} &  $1.820\pm 0.014$ &  $1.810\pm 0.016$ &  $2.248\pm 0.042$&  $2.020\pm 0.018$&  $1.620\pm 0.018$&  $2.208\pm 0.100$&  $1.944\pm 0.028$\\ 
           \bm{$q_{\rm m}$} &  $0.768\pm 0.004$  &  $ 0.904\pm 0.011$&  $ 0.885\pm 0.010$&   $ 0.703\pm 0.005$ &  $ 0.434\pm 0.006$&  $0.617\pm 0.045$&  $ 0.837\pm  0.010$ \\ 
           \bm{$ R_{\rm Eins}$} (arcsec)&  $3.271\pm 0.006$&  $ 1.902\pm 0.003$&  $ 1.392\pm 0.004$&  $ 3.586\pm 0.009$&  $ 2.875\pm 0.047$&   $ 2.333\pm 0.044$&  $ 1.494\pm 0.003 $\\ 
           \bm{$\varphi_{\rm m}$} (rad, +x) &  $1.587\pm 0.009$&   $ 0.268\pm 0.065$&  $ 1.719 \pm 0.045$ &   $ 2.249\pm 0.006$ &  $ 0.129\pm 0.002$&  $ 3.029\pm 0.023$&  $ 3.134\pm 0.024$\\ 
           \hline
            & \multicolumn{6}{l}{External shear} &\\
           \hline
           \bm{$\gamma_{\rm ext}$} &  $0.086\pm 0.003$&  $ 0.069\pm 0.004$&  $ 0.046\pm 0.003$&  $ 0.037\pm 0.002$&  $ 0.069\pm 0.005$&  $ 0.028\pm 0.011$&  $ 0.061\pm 0.005$\\ 
           \bm{$\varphi_{\rm ext}$} (rad, +x) &   $2.557\pm 0.012$&  $ 2.059\pm 0.034$&  $ 2.322\pm 0.027$ &  $ 2.208\pm 0.014$&  $ 0.436\pm 0.034$ &  $ 1.690\pm 0.13$ &   $ 1.559\pm0.018 $\\ 
           \hline
            & \multicolumn{6}{l}{S\'ersic profile 1} &\\
           \hline
           \bm{$q_{\rm L, 1}$} & $0.756\pm 0.002$& $ 0.766\pm 0.004$&  $ 0.994\pm 0.005$&  $ 0.698\pm 0.005$& $ 0.491\pm 0.003$& $ 0.711\pm 0.004$& $ 0.311\pm 0.006$\\  
           \bm{$R_{\rm eff, 1}$} (arcsec)&  $4.999\pm 0.032$&  $ 0.336\pm 0.010$&  $ 1.477\pm 0.014$&  $ 4.010\pm 0.114$&  $ 3.332\pm 0.037$ &   $ 2.741\pm 0.059$&   $ 2.603\pm 0.001$\\ 
           \bm{$\varphi_{\rm L, 1}$} (rad, +x)&  $1.323\pm 0.004$&  $ 0.556\pm 0.009$ &  $ 2.045\pm 0.900$&  $ 2.021\pm 0.007$ &  $ 0.173\pm 0.003$ &  $ 2.916\pm 0.007$&  $ 1.423\pm 0.013$\\ 
           \bm{$n_{1}$} & 1.970$\pm$0.057&  1.604$\pm$0.037&  0.598$\pm$0.021&  3.422$\pm$ 0.242&  1.681$\pm$0.03&  3.232$\pm$0.158&  2.718$\pm$0.001\\  
           \bm{$A_{1}$} ($e-/s$) & 0.226$\pm$0.003& 7.694$\pm$0.257 & 0.417$\pm$0.012 & 0.099$\pm$0.005& 0.206$\pm$0.004& 0.178$\pm$0.007 & 0.677$\pm$0.034\\ 
           \hline
          & \multicolumn{6}{l}{S\'ersic profile 2} & \\
           \hline
            \bm{$q_{\rm L, 2}$} & 0.846$\pm$0.002& 0.776$\pm$0.002 & 0.956$\pm$0.005& 0.921$\pm$0.013& 0.967$\pm$0.011& 0.899$\pm$0.014& 0.960$\pm$0.010 \\ 
            \bm{$R_{\rm eff, 2}$} (arcsec) & 0.451$\pm$0.007 & 2.262$\pm$0.021&  0.267$\pm$0.007& 0.278$\pm$0.010& 0.287$\pm$0.007& 0.259$\pm$0.007& 0.234$\pm$0.006\\  
            \bm{$\varphi_{\rm L, 2}$} (rad, +x)& 1.957$\pm$0.012& 0.553$\pm$0.006& 2.000$\pm$0.061& 2.459$\pm$0.100& 2.593$\pm$0.232&  0.059$\pm$0.073& 1.019$\pm$0.128 \\  
            \bm{$n_{2}$} & 1.681$\pm$0.020& 1.131$\pm$0.045&  1.760$\pm$0.055& 1.804$\pm$0.063& 1.613$\pm$0.042& 1.253$\pm$0.059& 1.196$\pm$0.048\\  
            \bm{$A_{2}$} ($e-/s$) & 7.149$\pm$0.107& 0.467$\pm$0.01& 6.703$\pm$0.256& 3.823$\pm$0.123& 2.789$\pm$0.082 & 3.705$\pm$0.156& 3.425$\pm$0.129\\ 
            \hline
             & \multicolumn{6}{l}{Single S\'ersic fit} &\\
            \hline
            \bm{$q_{\rm L}$}& 0.814$\pm$0.001& 0.766$\pm$0.001&0.965$\pm$0.002 & 0.766$\pm$0.002& 0.553$\pm$0.001&0.744$\pm$0.002 &0.965$\pm$0.004 \\
            \bm{$R_{\rm eff}$} (arcsec)& 6.498$\pm$0.048& 2.409$\pm$0.013& 1.458$\pm$0.015&3.042$\pm$0.039& 3.171$\pm$0.025 &2.213$\pm$0.019&2.100$\pm$0.042\\
            \bm{$\varphi_{\rm L}$} (rad, +x)&1.478$\pm$0.003& 0.549$\pm$0.003& 2.020$\pm$0.035&2.069$\pm$0.005&0.164$\pm$0.002&2.939$\pm$0.004& 0.607$\pm$0.064\\
            \bm{$n$} & 6.860$\pm$0.023&4.952$\pm$0.019& 5.740$\pm$0.044& 6.196$\pm$0.044& 3.866$\pm$0.026& 5.091$\pm$0.032& 4.849$\pm$0.058\\
            \bm{$A$} ($e-/s$)& 0.120$\pm$0.002& 0.397$\pm$0.004& 0.406$\pm$0.008&0.145$\pm$0.003&0.187$\pm$0.003& 0.252$\pm$0.004&0.125$\pm$0.004\\
            \hline
            & \multicolumn{6}{l}{Circularized half-light radius of lens galaxy (in arcsec)} &\\
            \hline
             \bm{$R_{\rm half, gal, eq}$} & 2.191$\pm$0.079 & 0.881$\pm$0.044 & 0.675$\pm$0.042 & 1.736$\pm$0.263 & 1.065$\pm$0.045 & 1.259$\pm$0.142 & 0.964$\pm$0.082 \\
            \hline
    \end{tabular}
    \caption{Parameters for the SPEMD power-law lens mass distribution, external shear, and lens light profile modeled using two S\'ersic functions. We also provide parameters for the single S\'ersic fit  to the lens light.
Here, $\gamma^{\rm lens}$ is magnitude of the slope of power-law density profile ($\rho \propto r^{-\gamma^{\rm lens}}$), $q_{\rm m}$ is lens mass axis ratio, $ R_{\rm Eins}$ is projected Einstein radius along major-axis, $  \varphi_{\rm m}$ is position angle in radians measured anti-clockwise from the positive x-axis, $\gamma_{\rm ext}$ is the external shear magnitude, $\varphi_{\rm ext}$ is the external shear angle in radians measured anti-clockwise from the positive x-axis. 
These are followed by S\'ersic profile parameters, where $q_{\rm L}$ is the axis ratio of lens light, $R_{\rm eff}$ is the effective half-light radius of the S\'ersic profile along semi-major axis, $\varphi_{\rm L}$ is lens light position angle,  $n$ is S\'ersic index, and $A$ is the amplitude representing flux count rate at $R_{\rm eff}$. $R_{\rm half, gal, eq}$ is the overall circularized half-light radius of the lens galaxy.
Subscripts 1 and 2 denote the two S\'ersic components of the total light profile. Mass and light centroids for \agel0537, \agel2335, \agel0142, and \agel0102  were linked during lens modeling. Mass and light centroids for \agel2158, \agel2336, and \agel1507  were not linked during modeling; however, both are found to be consistent within observed image pixel resolution ($\pm 0.08\arcsec$).}
    \label{table:lens parameters}
\end{table}

\section{L\&D versus Lensing-only density profile slopes}
\label{appendix:LnD_vs_LensingOnly}

\begin{figure}
    \centering
    \includegraphics[clip=true,trim= 02mm 15mm 01mm 15mm,width=  0.5\textwidth]{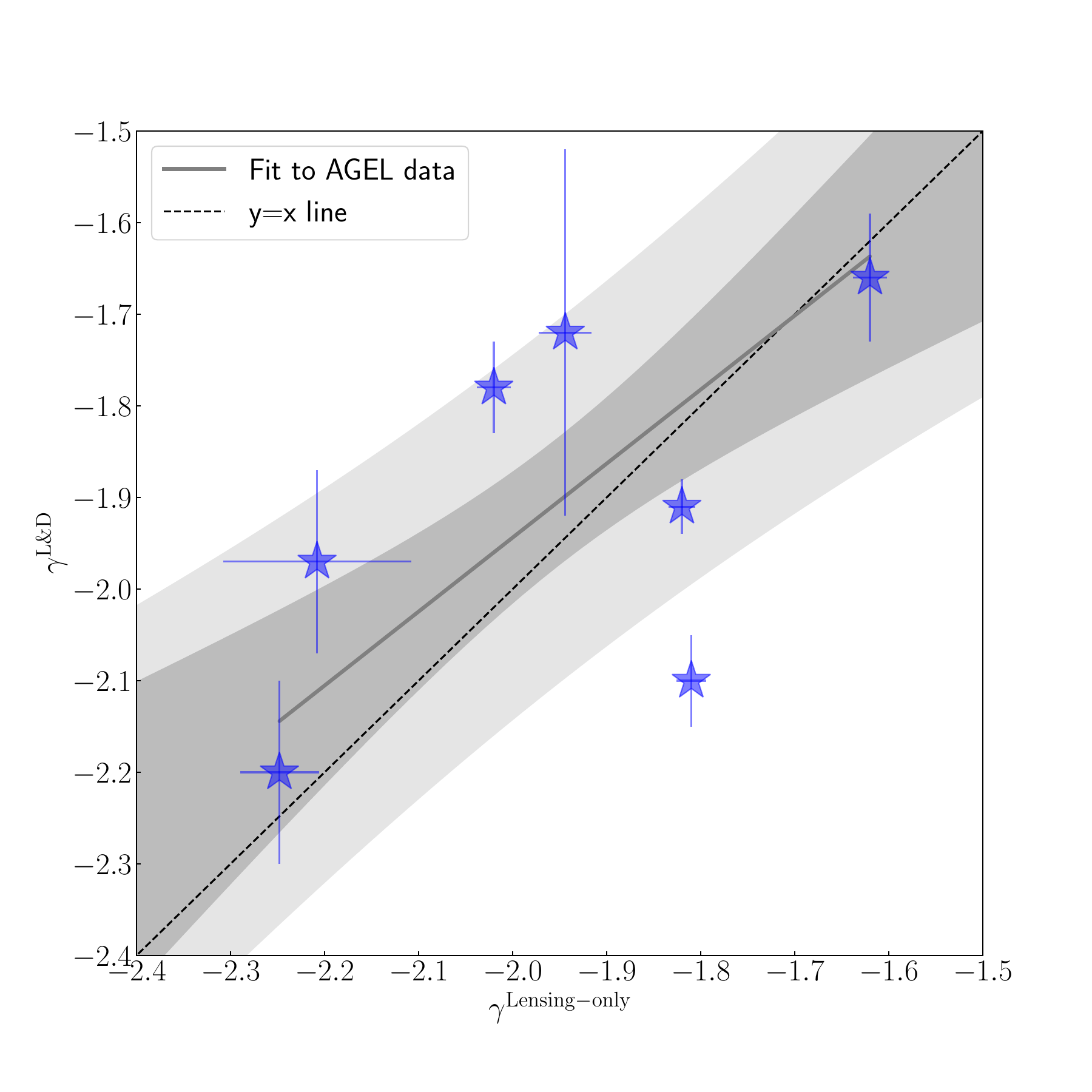}
    \caption{One-to-one comparison between total density profile slope of lenses in \agel \ sample using the two methods: lensing-only and L\&D. Gray line is a fit to the data with dark gray shades showing $1\sigma$ uncertainty in line slope and intercept, and light shaded region shows $1\sigma$ scatter about the fit. }
    \label{fig:gamma-gamma}
\end{figure}

\begin{figure}
    \centering
    \includegraphics[clip=true,trim= 02mm 02mm 01mm 02mm,width= \textwidth]{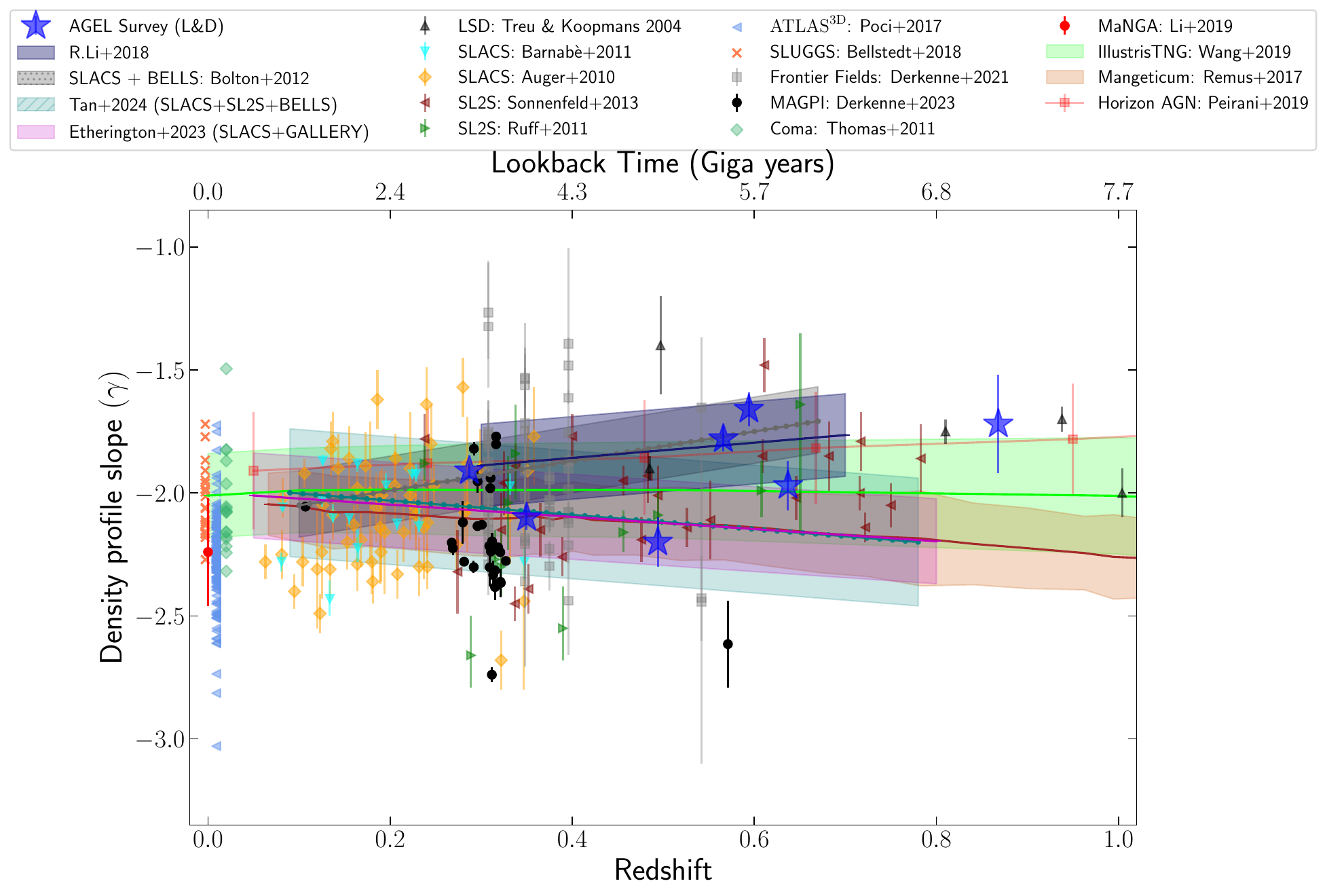}
    \caption{Same as Figure \ref{fig:gam-z all} but now representing L\&D slopes for the \agel \ pilot sample marked with blue stars.}
    \label{fig:gamma-z_All_LnD}
\end{figure}

Figure \ref{fig:gamma-gamma} compares the lensing-only density profile slopes with the slopes obtained using joint lensing and dynamical analysis.
Figure \ref{fig:gamma-gamma} also shows a bisector fit, performed using the regression routine from  \citet{Akritas:Bershady:1996}, to quantify the correlation between the two measurements.
The $\gamma$--$z$ diagram with the L\&D slopes for the \agel \ pilot sample is shown in Figure \ref{fig:gamma-z_All_LnD}. See Section \ref{subsec: lensing} of the main paper for more details.

\end{document}